\documentclass[12pt]{article}
\usepackage{amssymb,amsmath,epsfig}

\begin{document}
\title{\bf Cosmology of Holographic and New Agegraphic $f(R,T)$ Models}

\author{Muhammad SHARIF \thanks{msharif.math@pu.edu.pk} and Muhammad
ZUBAIR
\thanks{mzubairkk@gmail.com}\\\\
Department of Mathematics, University of the Punjab,\\
Quaid-e-Azam Campus, Lahore-54590, Pakistan.}

\date{}

\maketitle

\begin{abstract}
We consider the $f(R,T)$ theory, where $R$ is the scalar curvature
and $T$ is the trace of energy-momentum tensor, as an effective
description for the holographic and new agegraphic dark energy and
reconstruct the corresponding $f(R,T)$ functions. In this study, we
concentrate on two particular models of $f(R,T)$ gravity namely,
$R+2A(T)$ and $B(R)+\lambda{T}$. We conclude that the derived
$f(R,T)$ models can represent phantom or quintessence regimes of the
universe which are compatible with the current observational data.
In addition, the conditions to preserve the generalized second law
of thermodynamics are established.
\end{abstract}

\section{Introduction}

Supernovae type Ia (SNeIa)$^{{1})}$ observations revealed the
expanding behavior of the universe. This fact has further been
affirmed by the observations of anisotropies in cosmic microwave
background (CMB)$^{{2})}$, large scale structure$^{{3})}$, baryon
acoustic oscillations$^{{4})}$ and weak lensing$^{{5})}$. A strange
type of energy component with prominent negative pressure identified
as \emph{dark energy} (DE) is used to explain the current cosmic
acceleration. The source and characteristics of DE are still a
complicated story as several models have been suggested in the
context of general relativity (GR) (for review see$^{{6})}$).

The most likely campaigner of DE is the cosmological constant or the
vacuum energy whose equation of state (EoS) parameter is fixed,
$\omega_{\Lambda}=-1$. The cosmological model that consists of
cosmological constant plus cold dark matter is entitled as
${\Lambda}CDM$ model, which appears to fit the observational data.
However, despite of its success, this model experiences two notable
cosmological problems namely, the ``fine tuning" problem and the
``cosmic coincidence" problem$^{{7})}$. Such issue primarily
originates because the vacuum energy is counted in the setting of
quantum field theory in Minkowski background. Nevertheless, it is
considerably accepted that at cosmological measures where the
quantum effects of gravity may be reported, the preceding sketch of
vacuum energy would not sustain.

The accurate measurement of the vacuum energy may be indicated by
comprehensive quantum theory of gravity. Though, we are lacking such
a profound theory, it is possible to investigate the nature of DE
corresponding to some principles of quantum gravity. In particular,
the holographic principle$^{{8})}$ is a significant characteristic
that may play role to deal with cosmological and DE issues. Cohen et
al.$^{{9})}$ suggested a relation between the infrared (IR) and
ultraviolet (UV) cutoffs because of the limit made by the formation
of black hole, which adjusts up an upper bound for the vacuum energy
$L^3\rho_{\vartheta}\leqslant{L}M_p^2$, where $\rho_\vartheta$ is
the vacuum energy associated with the UV cutoff, $L$ is the IR
cutoff and $M_p $ is the reduced Planck mass. Li$^{{10})}$ proposed
the form of DE and suggested that the future event horizon is the
appropriate choice for IR cutoff which seems to agree with recent
measurements$^{{11})}$.

Introducing new ingredients of DE to the entire cosmic energy is the
one approach to explain the mystery of cosmic acceleration. Another
approach is based on modification of the Einstein-Hilbert action to
get alternative theories of gravity such as $f(R)$$^{{12})}$,
$f(\mathcal{T})$$^{{13})}$, where $\mathcal{T}$ is the the torsion
and $f(R,T)$ theory$^{{14})}$ etc. Harko et al.$^{{14})}$ introduced
$f(R,T)$ theory by generalizing $f(R)$ gravity and is established on
the coupling between matter and geometry. Recently, this theory has
gained attention and some worth mentioning results have been
explored$^{{15-20})}$.

Many authors$^{{21-27})}$ have discussed the cosmological
reconstruction of modified theories of gravity according to
holographic DE. Karami and Khaledian$^{{25})}$ reconstructed $f(R)$
models according to holographic and new agegraphic DE. Daouda et al.
$^{{26})}$ develped $f(\mathcal{T})$ model using holographic DE
which can imply unified scenario of dark matter with DE. Houndjo and
Piattella$^{{17})}$ numerically reconstructed the $f(R,T)$ models
which can represent the characteristics of holographic DE models. In
this work, we consider the holographic and new agegraphic DE models,
and reconstruct the corresponding $f(R,T)$ gravity as an equivalent
picture without utilizing any additional DE component. We also
investigate the generalized second law of thermodynamics (GSLT) on
the future event horizon and find out the necessary condition for
its validity.

The paper is arranged as follows. In the next section, we introduce
the general formulation of the field equations in $f(R,T)$ gravity.
Sections \textbf{3} and \textbf{4} provide the reconstruction of
$f(R,T)$ gravity according to holographic and new agegraphic DE
respectively. In section \textbf{5}, the validity of GSLT is
investigated and the last section concludes our results.

\section{$f(R,T)$ Gravity: General Formalism}

The $f(R,T)$ gravity is an appealing modification to the
Einstein-Hilbert action by setting an arbitrary function of scalar
curvature $R$ and trace of the energy-momentum tensor $T$. The
action for this theory is defined as$^{{14})}$
\begin{equation}\label{1}
\mathcal{I}=\int{dx^4\sqrt{-g}\left[\frac{M_p^2}{2}f(R,T)+\mathcal{L}_{(M)}\right]},
\end{equation}
where $M_p^{-2}=8\pi{G}$ and $\hbar=c=1$. The energy-momentum tensor
of matter component is determined as$^{{28})}$
\begin{equation}\label{2}
T^{(M)}_{\alpha\beta}=-\frac{2}{\sqrt{-g}}\frac{\delta(\sqrt{-g}
{\mathcal{\mathcal{L}}_{(M)}})}{\delta{g^{\alpha\beta}}}.
\end{equation}
The correspong field equations are found through the variation of
(\ref{1}) with respect to the metric tensor
\begin{eqnarray}\label{3}
&&R_{\alpha\beta}f_{R}(R,T)-\frac{1}{2}g_{\alpha\beta}f(R,T)+(g_{\alpha\beta}
{\Box}-{\nabla}_{\alpha}{\nabla}_{\beta})f_{R}(R,T)\nonumber\\&=&M_p^{-2}
T^{(M)}_{\alpha\beta}-f_{T}(R,T)T^{(M)}_{\alpha\beta}-f_{T}(R,T)\Theta_{\alpha\beta},
\end{eqnarray}
where
$f_{R}={\partial}f/{\partial}R,~f_{T}={\partial}f/{\partial}T,~
{\Box}={\nabla}_{\alpha}{\nabla}^{\beta};~{\nabla}_{\alpha}$ is the
covariant derivative linked with the Levi-Civita connection symbol
and $\Theta_{\alpha\beta}$ is defined by
\begin{equation}\label{4}
\Theta_{\alpha\beta}=\frac{g^{\mu\nu}{\delta}T_{\mu\nu}^{(M)}}
{{\delta}g^{\alpha\beta}}=-2T_{\alpha\beta}^{(M)}+g_{\alpha\beta}\mathcal{L}_M
-2g^{\mu\nu}\frac{\partial^2\mathcal{L}_M}{{\partial}
g^{\alpha\beta}{\partial}g^{\mu\nu}}.
\end{equation}
The matter content is assumed to be perfect fluid so that
\begin{equation*}
T_{\alpha\beta}^{(M)}=({\rho}_M+p_M)u_{\alpha}u_{\beta}-p_{M}g_{\alpha\beta},
\end{equation*}
where $u_{\alpha}$ is the four velocity which satisfies
$u_{\alpha}u^{\alpha}=1$, $\rho_M$ and $p_M$ are the energy density
and pressure of the fluid, respectively.  The matter Lagrangian can
be assumed as $\mathcal{L}_{M}=-p_M$, so that $\Theta_{\alpha\beta}$
becomes
\begin{equation}\label{5}
\Theta_{\alpha\beta}=-2T_{\alpha\beta}^{(M)}-p_{M}g_{\alpha\beta}.
\end{equation}

We assume the $f(R,T)$ model as $f(R,T)=f_1(R)+f_2(T)$, where
$f_{1}$ and $f_{2}$ are arbitrary functions of $R$ and $T$,
respectively. Thus the field equation (\ref{3}) becomes
\begin{eqnarray}\label{6}
&&R_{\alpha\beta}f_{1R}-\frac{1}{2}g_{\alpha\beta}f_{1}+(g_{\alpha\beta}
{\Box}-{\nabla}_{\alpha}{\nabla}_{\beta})f_{1R}=M^{-2}_p
T_{\alpha\beta}^{(M)}+T_{\alpha\beta}^{(M)}f_{2T}\nonumber\\
&+&[pf_{2T}+\frac{1}{2}f_{2}]g_{\alpha\beta},
\end{eqnarray}
which can be reproduced as an effective Einstein field equation,
\emph{i.e.},
\begin{equation}\label{7}
R_{\alpha\beta}-\frac{1}{2}Rg_{\alpha\beta}
=\tilde{M}^{-2}_pT_{\alpha\beta}^{EFF},
\end{equation}
where $\tilde{M}^{-2}_p=(M^{-2}_p+f_{2T})/f_{1R}$ and
\begin{eqnarray*}
{T}_{\alpha\beta}^{EFF}&=&{T}_{\alpha\beta}^{(M)}+\frac{\tilde{M}^{2}_p}{f_{R}}\left[\frac{1}{2}(f_1+f_2+
2p_Mf_{2T}-Rf_{1R})g_{\alpha\beta}+({\nabla}_{\alpha}{\nabla}_{\beta}-g_{\alpha\beta}{\Box})f_{1R}\right].
\end{eqnarray*}
Now, we formulate the field equations of $f(R,T)$ models for
particular choices of $f_1$ and $f_2$.

\subsection{$f(R,T)=R+2A(T)$ Gravity}

We propose a particular case with $f_1(R)=R$ and $f_2(T)=2A(T)$.
Such model appears to be interesting and has been widely studied in
literature$^{{16-19})}$. Accordingly, the field equations are
obtained as follows
\begin{equation}\nonumber
R_{\alpha\beta}-\frac{1}{2}Rg_{\alpha\beta}=(M^{-2}_p+2A_T(T))T^{(M)}_{\alpha\beta}
+(2p_MA_T(T)+A(T))g_{\alpha\beta}.
\end{equation}
The line element of spatially flat FRW spacetime is given by
\begin{equation}\label{8}
ds^{2}=dt^2-a^2(t)d\textbf{x}^2,
\end{equation}
where $a(t)$ is the scale factor and $d\textbf{x}^2$ comprises the
spatial part of the metric. In this background, the above field
equations can be represented as
\begin{eqnarray}\label{9}
3M^{2}_pH^2&=&\rho_M+\rho_{dc}, \\\label{10}
-M^{2}_p(2\dot{H}+3H^2)&=&p_{M}+p_{dc},
\end{eqnarray}
where $H=\dot{a}/{a}$ is the Hubble parameter and dot represents
differentiation with respect to time. The energy density
($\rho_{dc}$) and pressure ($p_{dc}$) of \emph{dark energy
components} are obtained as
\begin{eqnarray}\label{11}
\rho_{dc}&=&M^{2}_p[2(\rho_M+p_M)A_T(T)+A(T)], \\\label{12}
p_{dc}&=&-M^{2}_pA(T).
\end{eqnarray}
The corresponding EoS parameter is
\begin{equation}\label{13}
\omega_{dc}=\frac{-A(T)}{2(\rho_M+p_M)A_T(T)+A(T)}.
\end{equation}

\subsection{$f(R,T)=B(R)+{\lambda}T$ Gravity}

Let us consider a more complicated case choosing $f_1(R)=B(R)$ and
$f_2(T)={\lambda}T$$^{{18-20})}$, ${\lambda}T$ can be considered as
correction term to $f(R)$ gravity. For this model, the field
equation (\ref{7}) can be represented as
\begin{equation}\label{14}
\tilde{M}^{2}_p\left(R_{\alpha\beta}-\frac{1}{2}Rg_{\alpha\beta}\right)
=T^{(M)}_{\alpha\beta}+T^{(dc)}_{\alpha\beta},
\end{equation}
where $\tilde{M}^{-2}_p=(M^{-2}_p+\lambda)/B_R$ and
\begin{eqnarray*}
{T}_{\alpha\beta}^{(dc)}&=&\frac{\tilde{M}^{2}_p}{B_{R}}\left[\frac{\lambda}{2}(\rho_M-p_M)g_{\alpha\beta}+
\frac{1}{2}(B-RB_{R})g_{\alpha\beta}+({\nabla}_{\alpha}{\nabla}_{\beta}-g_{\alpha\beta}{\Box})B_{R}\right].
\end{eqnarray*}
For the choice of pressureless matter, Eq.(\ref{14}) can be
rewritten in terms of FRW equations (\ref{9}) and (\ref{10}), where
\begin{eqnarray}\label{15}
\rho_{dc}&=&\tilde{M}^{2}_p\left[\frac{3\lambda}{2}\rho_M+\frac{1}{2}(B-RB_{R})
-3H\dot{R}B_{RR}+3H^2(1-B_R)\right],\\\nonumber
p_{dc}&=&\tilde{M}^{2}_p\left[-\frac{\lambda}{2}\rho_M+\frac{1}{2}(RB_{R}-B)+
(\ddot{R}+2H\dot{R})B_{RR}+\dot{R}^2B_{RRR}\right.\\\label{16}&-&\left.
(2\dot{H}+3H^2)(1-B_R)\right].
\end{eqnarray}
Using Eqs.(15) and (\ref{16}), we can develop the evolution equation
for $B(R)$ as
\begin{eqnarray}\label{17}
\dot{R}^2B_{RRR}+(\ddot{R}-H\dot{R})B_{RR}+2\dot{H}(B_R-1)+
{\lambda}\rho_M-M^{-2}_p(1+\omega_{dc})\rho_{dc}=0.
\end{eqnarray}
This represents a third order differential equation in $B(R)$. In
sections \textbf{3} and \textbf{4}, we reconstruct the $f(R,T)$
models for holographic DE (HDE) and new agegraphic DE (NADE) as
follows.

\section{Reconstruction from Holographic Dark Energy}

According to holographic principle$^{{8})}$, the HDE density is
given by$^{{9})}$
\begin{equation}\label{18}
\rho_{\vartheta}=\frac{3e^2M^{2}_p}{L^2},
\end{equation}
where $e$ is a constant. The IR cutoff $L$ (future event horizon) is
defined as$^{{10})}$
\begin{equation*}
L=R_{\hat{E}}=a(t)\int^{\infty}_{t}{\frac{d\hat{t}}{a(\hat{t})}}
=a(t)\int^{\infty}_{a}{\frac{da'}{Ha'^{2}}} .
\end{equation*}
For the homogeneous and isotropic universe with spatially flat
geometry, comprising matter component and HDE, the Friedmann
equation reads
\begin{equation}\label{19}
3M^{2}_pH^2=\rho_M+\rho_{\vartheta},
\end{equation}
where $\rho_M=\rho_{M0}(1+z)^3$ from the energy conservation
equation of matter. By introducing critical energy density
$\rho_{cri}=3M^{2}_pH^2$ and dimensionless DE
$\Omega_{\vartheta}=\frac{\rho_{\vartheta}}{\rho_{cri}}$, we obtain
\begin{equation}\label{20}
\dot{R}_{\hat{E}}=HR_{\hat{E}}-1=\frac{e}{\sqrt{\Omega_{\vartheta}}}-1.
\end{equation}
The HDE satisfies the conservation law
\begin{eqnarray}\label{21}
\dot{\rho}_{\vartheta}+3H\rho_{\vartheta}(1+\omega_{\vartheta})&=&0.
\end{eqnarray}
Using Eqs.(\ref{18}) and (\ref{20}), the time derivative of HDE
reads as
\begin{equation}\label{22}
\dot{\rho}_{\vartheta}=\frac{-2}{R_h}\left(\frac{e}{\sqrt{\Omega_{\vartheta}}}-1\right)\rho_{\vartheta}.
\end{equation}
Combining Eqs.(\ref{21}) and (\ref{22}), the EoS parameter of HDE
becomes
\begin{equation}\label{23}
\omega_{\vartheta}=-\frac{1}{3}\left(1+\frac{2\sqrt{\Omega_{\vartheta}}}{e}\right).
\end{equation}

It can be seen that when $\Omega_{\vartheta}\longrightarrow1$ in the
future (\emph{i.e.}, the HDE dominates the contents of the
universe), for $e>1$, we have  $\omega_{\vartheta}>-1$ which depicts
quintessence era such that the universe escapes from entering the de
Sitter and Big Rip phases. For $e=1$, it represents the de Sitter
universe and if $e<1$, it may end up with phantom phase and behaves
as quintom era because EoS parameter intersects the cosmological
constant boundary (the phantom divide) throughout evolution. Hence,
the parameter $e$ plays a significant character in determining the
evolutionary paradigm of HDE as well as ultimate fate of the
universe. The HDE has been constrained from observations of SNeIa,
CMB and galaxy clusters, the best fit favors $e<1$, although $e>1$
is also compatible with the data in one-sigma error range$^{{11})}$.

Now we reconstruct the HDE $f(R,T)$ models by considering two
particular actions of $f(R,T)$ Lagrangian. {\begin{itemize}
\item {$R+2A(T)$}
\end{itemize}}
Comparing EoS parameter of \emph{dark energy components}
$\omega_{dc}$$^{{13})}$ for the above model with that of HDE, one
obtains
\begin{equation}\label{24}
\frac{A(T)}{2(\rho_M+p_M)A_T(T)+A(T)}=\frac{1}{3}\left(1+\frac{2\sqrt{\Omega_{\vartheta}}}{e}\right).
\end{equation}
For the standard model (\ref{19}), we consider the pressureless
matter so that Eq.(\ref{24}) is manipulated as
\begin{equation}\label{25}
TA_T-\frac{e-\sqrt{\Omega_{\vartheta}}}{e+2\sqrt{\Omega_{\vartheta}}}A=0.
\end{equation}
This is the first order differential equation. For constant
$\Omega_{\vartheta}$, its solution is of the form
$$A(T)\propto{T}^{\frac{e-\sqrt{\Omega_{\vartheta}}}{e+2\sqrt{\Omega_{\vartheta}}}}.$$
We are interested to determine the $A(T)$ model coming from HDE.
Also, for a given $a(t)$, the $f(R,T)$ gravity can be reconstructed
corresponding to any DE model. The Hubble parameter $H$ is assumed
to be
\begin{equation}\label{26}
H(t)=m(t_p-t)^{-\epsilon},
\end{equation}
where $m$ and $\epsilon$ are positive constants and $t<t_p,~t_p$ is
the probable time when finite-time future singularity may appear.
$H(t)$  given by (\ref{26}) specifies two type of singularities,
type \textbf{I} (``Big rip singularity") and type \textbf{III} which
can occur for $\epsilon\geqslant1$ and $0<\epsilon<1$ respectively.
One can find details of the classification of finite-time
singularities in literature$^{{29})}$.

We look at the elementary case by choosing $\epsilon=1$ so that
$a(t)=a_0(t_p-t)^{-m},~a_0>0$ representing the phantom phase of the
universe which may result in Big rip singularity within finite time
$(t\rightarrow{t}_p)$. For this model, the future event horizon
$R_{\hat{E}}$ and $\Omega_{\vartheta}$ are obtained as
\begin{equation}\label{27}
R_{\hat{E}}=\frac{t_p-t}{m+1}, \quad
\sqrt{\Omega_{\vartheta}}=\frac{e(m+1)}{m}.
\end{equation}
Consequently, the solution of Eq.(\ref{25}) yields
\begin{equation}\label{28} A(T)=C{T}^{K},
\end{equation}
and the corresponding $f(R,T)$ HDE model is
\begin{equation}\label{29}
f(R,T)=R+2C{T}^{K},
\end{equation}
where $K=-1/(3m+2)$ is a constant depending on $m$ and $C$ is the
integration constant. To find the constant $C$, we need to develop
initial condition on $A(T)$. The Friedmann equation (\ref{9})
evaluated at $t=t_0$ yields
\begin{equation}\label{30}
[1+2A_T(T_0)]\Omega_{M0}+\frac{A(T_0)}{3H_0^2}=1.
\end{equation}
Manipulating Eqs.(\ref{25}) and (\ref{30}) at present time, it
follows that
\begin{equation}\label{31}
A(T_0)=3H_0^2{\Omega}_{\vartheta0}\left(1+2\frac{e-\sqrt{\Omega_{\vartheta0}}}
{e+2\sqrt{\Omega_{\vartheta0}}}\right)^{-1}.
\end{equation}
Applying initial condition (\ref{31}), the constant $C$ is
determined as
\begin{equation}\label{32}
C=3H_0^2{\Omega}_{\vartheta0}T_0^{-K}\left(1+2\frac{e-\sqrt{\Omega_{\vartheta0}}}
{e+2\sqrt{\Omega_{\vartheta0}}}\right)^{-1}.
\end{equation}
Hence, the explicit function of $f(R,T)$ is given by
\begin{equation}\label{33}
f(R,T)=R+6H_0^2{\Omega}_{\vartheta0}T_0^{-K}\left(1+2\frac{e-\sqrt{\Omega_{\vartheta0}}}
{e+2\sqrt{\Omega_{\vartheta0}}}\right)^{-1}T^K.
\end{equation}
\begin{figure}
\centering \epsfig{file=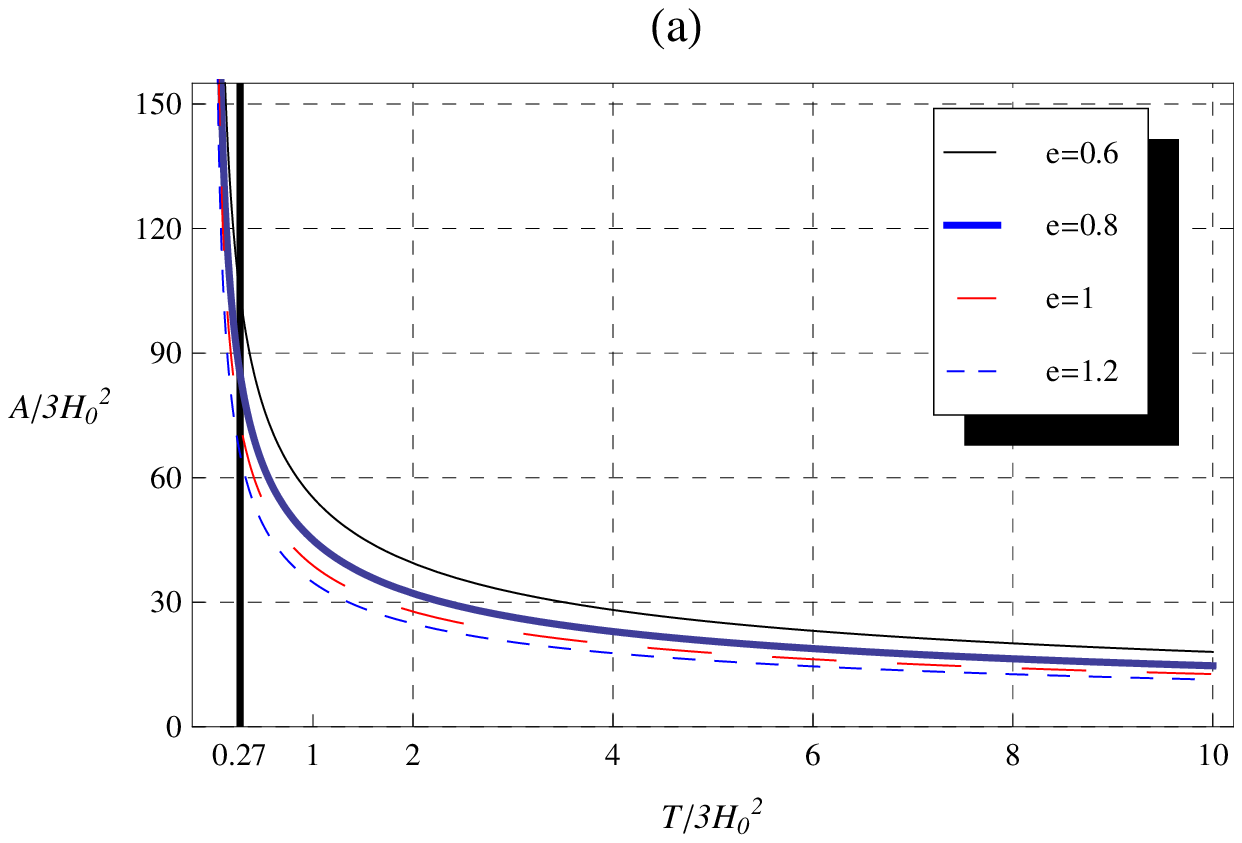, width=.495\linewidth,
height=2.2in} \epsfig{file=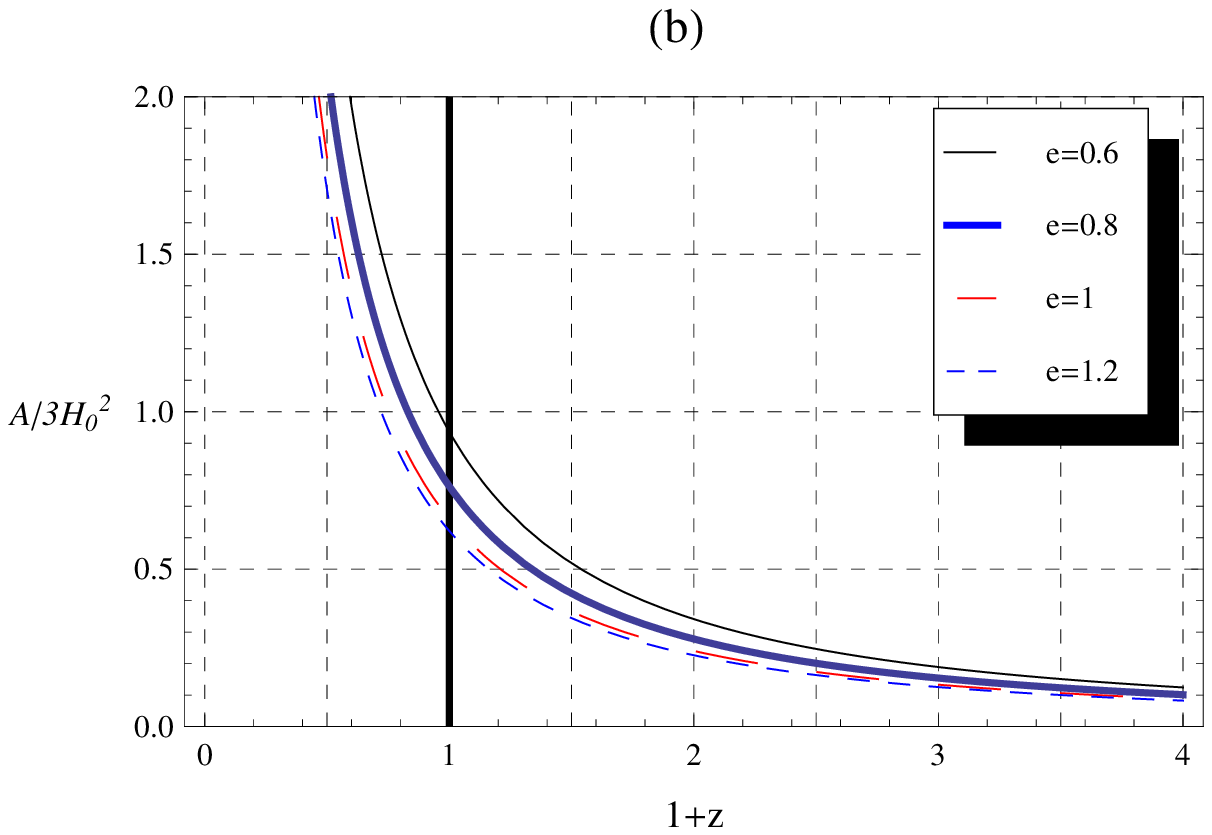, width=.495\linewidth,
height=2.2in} \caption{(Colour online) Evolution of $A(T)$ in HDE
(a) versus $T$ (b) versus $z$ for different values of parameter $e$.
Black thick line represents the current value of $T=T_0$. We set
$\Omega_{M0}=0.27$, $\Omega_{\vartheta0}=1-\Omega_{M0}$ and
$H_0=74$.}
\end{figure}
\begin{figure}
\centering \epsfig{file=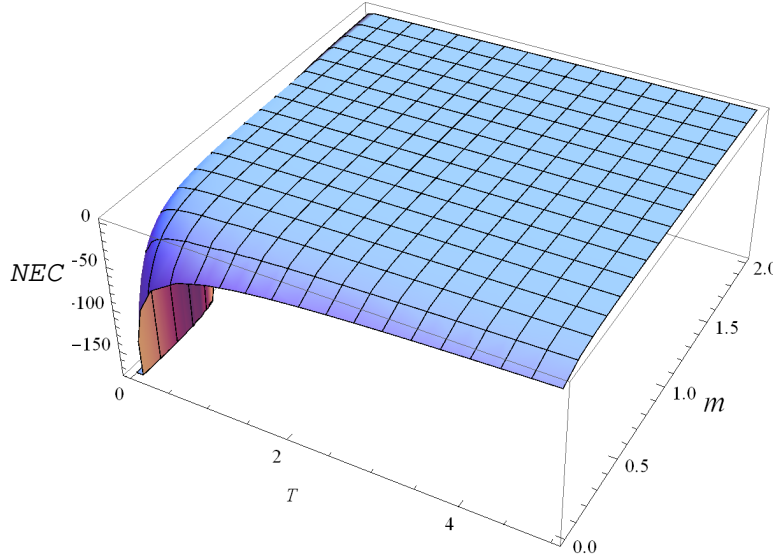}\caption{(Colour online) Evolution
of NEC in HDE versus $T$ and $m$.}
\end{figure}

In this representation, we normalize $A(T)$ and $T$ to $3H_0^2$ and
set $\Omega_{M0}=0.27$ and $e=0.6, 0.8, 1, 1.2$. The function $A(T)$
is plotted against $T$ and $z$ in Figure \textbf{1}. The difference
among the values of $e$ is apparent for earlier times of the
universe which vanishes in late times. Figure \textbf{1(b)} shows
the evolution in terms of redshift and here variation in curves is
evident in future evolution for different values of $e$. The
function $A(T)$ satisfies the EoS parameter
$\omega_{dc}=-1-\frac{2}{3m}$ which depicts the phantom era of DE.
Figure \textbf{2} clearly shows that for this $A(T)$ model, null
energy condition (NEC) is violated and hence accelerated expansion
of the universe is achievable. Here, NEC would violate even if one
increases the value of $m$ which is in agreement with EoS parameter
$\omega_{dc}$ for this model. {\begin{itemize}
\item {$B(R)+{\lambda}T$}
\end{itemize}}
Here, we reconstruct the function $B(R)$ in the setting of HDE. For
the choice of Hubble parameter $H(t)=\frac{m}{t_p-t}$, the future
event horizon and matter energy density can be rewritten in terms of
the Ricci scalar as
\begin{eqnarray}\label{34}
R_{\hat{E}}=\frac{1}{m+1}\sqrt{\frac{6m(2m+1)}{R}}, \quad
\rho_M=\frac{M^{2}_p[e^2(m+1)^2-m^2]}{2m(2m+1)}R.
\end{eqnarray}
Using Eqs.(\ref{18}) and (\ref{23}), one can get
\begin{eqnarray}\label{35}
(1+\omega_{\vartheta})\rho_\vartheta=\frac{M^{2}_pe^2(m+1)^2}{3m^2(2m+1)}R.
\end{eqnarray}
Substituting Eqs.(\ref{34}) and (\ref{35}) in Eq.(\ref{17}) and
solving, it follows that
\begin{equation}\label{36}
B(R)=\mu_{-}R^{\jmath_{-}}C_1+\mu_{+}R^{\jmath_{+}}C_2+\gamma{R}+C_3,
\end{equation}
where
\begin{eqnarray*}
\jmath_{\pm}&=&\frac{1}{4}\left[3+m\pm\sqrt{m^2-10m+1}\right], \quad
\quad \mu_{\pm}=\frac{1}{\jmath_{\pm}},\\\nonumber
\gamma&=&\frac{(2-3{\lambda}m)}{2m^2}\left[m^2-e^2(m+1)^2\right],
\end{eqnarray*}
$C_1,~C_2$ and $C_3$ are constants.

Now, we define necessary initial conditions to determine the values
of constants. For this purpose, we make the same assumption as in
ref.$^{{17})}$. In particular, we choose the initial conditions
$(B_R)_{t=t_0}=1$ and $(B_{RR})_{t=t_0}=0$ which can be translated
as
\begin{eqnarray}\label{37}
\left(\frac{dB}{dt}\right)_{t=t_0}=\left(\frac{dR}{dt}\right)_{t=t_0},
\quad
\left(\frac{d^2B}{dt^2}\right)_{t=t_0}=\left(\frac{d^2R}{dt^2}\right)_{t=t_0}.
\end{eqnarray}
Evaluating Eqs.(\ref{9}) and (\ref{15}), at $t=t_0$ and solving with
respect to $B(R_0)$, we ultimately have
\begin{equation}\label{38}
B(t=t_0)=R_0+\beta, \quad
\beta=6H_0^2(1-\Omega_{M0}-\frac{3}{2}{\lambda}M^{2}_p\Omega_{M0}).
\end{equation}
Applying the above initial conditions to the solution (\ref{36}), it
follows that
\begin{equation}\label{39}
B(R)=C_{+}R^{\jmath_{+}}+C_{-}R^{\jmath_{-}}+\gamma{R}+\delta,
\end{equation}
where
\begin{eqnarray*}
C_{+}=\frac{(\gamma-1)(\jmath_{-}-1)}{\jmath_{+}(\jmath_{+}-\jmath_{-})R_0^{\jmath_{+}-1}},
\quad
C_{-}=\frac{(\gamma-1)(\jmath_{+}-1)}{\jmath_{-}(\jmath_{-}-\jmath_{+})R_0^{\jmath_{-}-1}},
\\\nonumber
\delta=\beta+(1-\gamma)R_0+\frac{1}{\jmath_+\jmath_-}(\gamma-1)(\jmath_++\jmath_--1)R_0.
\end{eqnarray*}
Consequently, the $f(R,T)$ model corresponding to HDE turns out to
be
\begin{equation}\label{40}
f(R,T)=C_{+}R^{\jmath_{+}}+C_{-}R^{\jmath_{-}}+\gamma{R}+\delta+{\lambda}T.
\end{equation}
\begin{figure}
\centering \epsfig{file=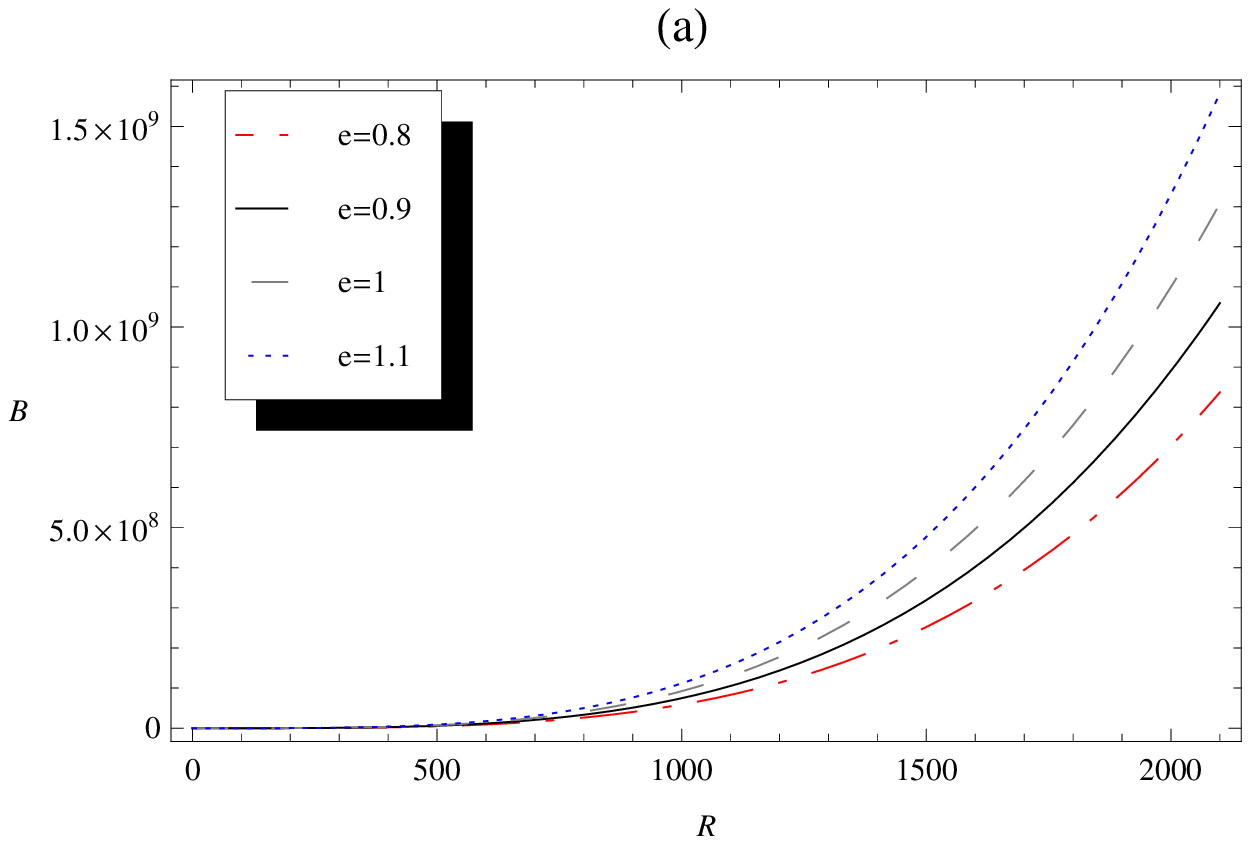, width=.495\linewidth,
height=2.2in} \epsfig{file=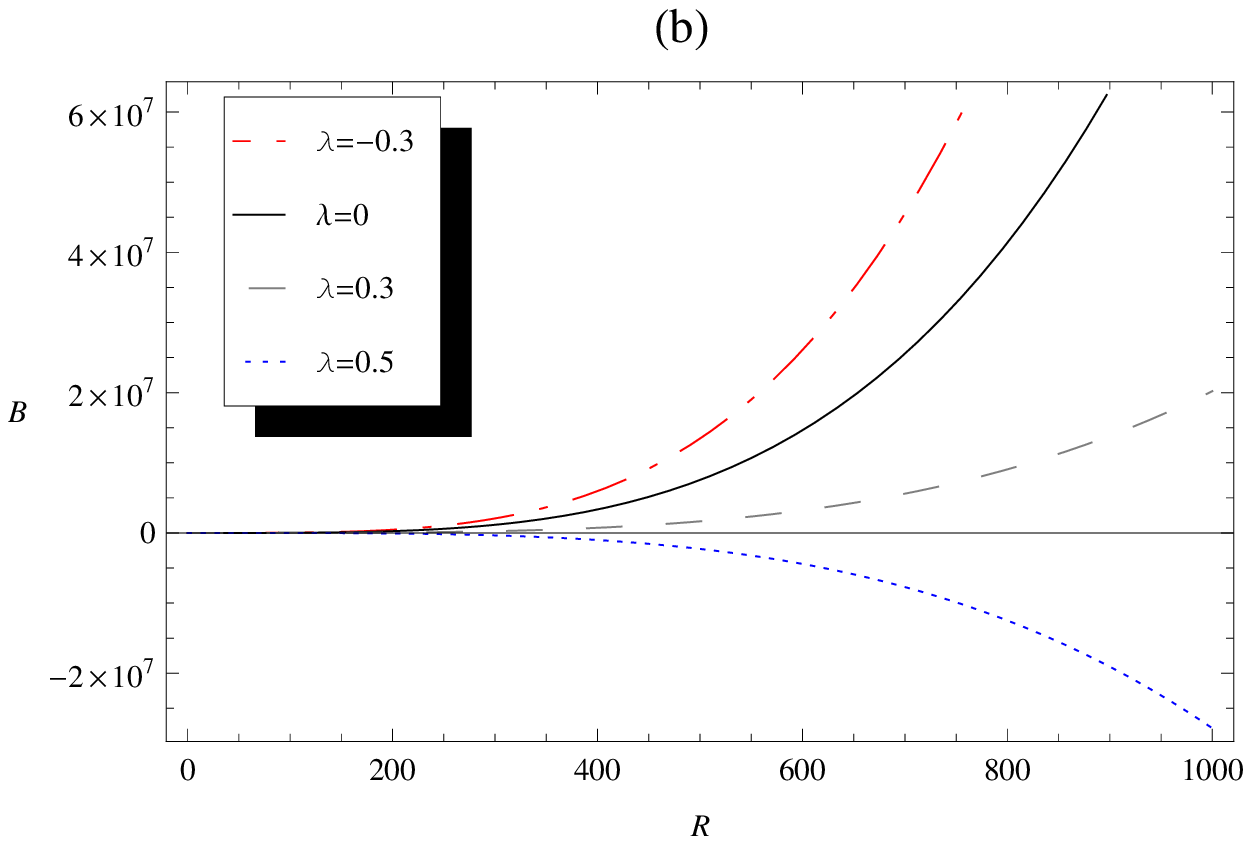, width=.495\linewidth,
height=2.2in} \caption{(Colour online) Evolution of $B(R)$ versus
$R$ in HDE for (a) different values of $e$ with $\lambda=0$ and (b)
different values of $\lambda$ with $e=1$.}
\end{figure}

We plot the function $B(R)$ against $R$ for different choices of
parameters $e$ and $\lambda$. In Figure \textbf{3(a)}, we fix
$\lambda=0$ (\emph{i.e.}, purely $f(R)$ gravity), which represents
the variation of $B$ for different values of parameter $e$. It is
obvious that curves become distinct for large $R$ and show
increasing behavior. The effect of coupling parameter $\lambda$ is
shown in Figure \textbf{3(b)} for $e=1$. We can see that non-zero
values of $\lambda$ modify the evolutionary nature of curves. We
have also represented these results in terms of redshift in Figure
\textbf{4}. These curves exhibit the future evolution of $B$ for
different values of parameters $e$ and $\lambda$.
\begin{figure}
\centering \epsfig{file=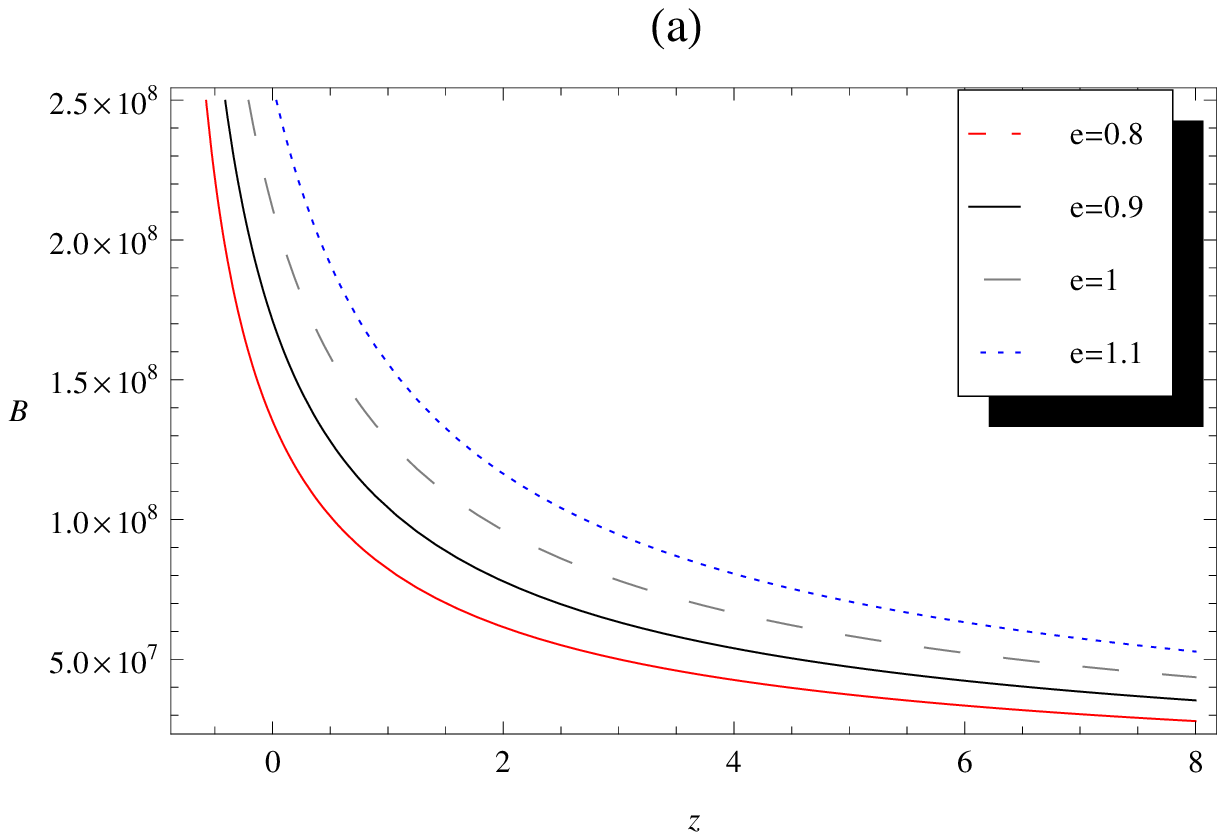, width=.495\linewidth,
height=2.2in} \epsfig{file=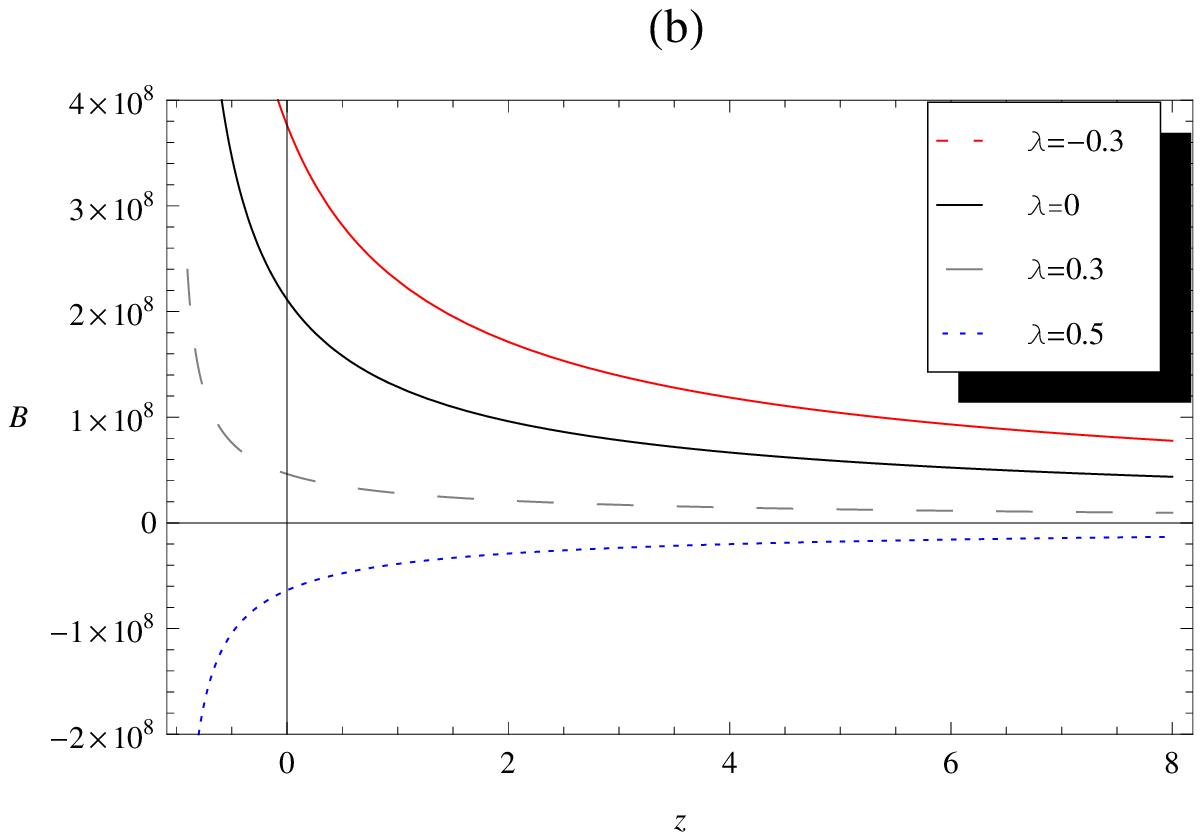, width=.495\linewidth,
height=2.2in} \caption{(Colour online) Evolution of $B(R)$ versus
$z$ in HDE for (a) different values of $e$ with $\lambda=0$ and (b)
different values of $\lambda$ with $e=1$.}
\end{figure}
\begin{figure}
\centering \epsfig{file=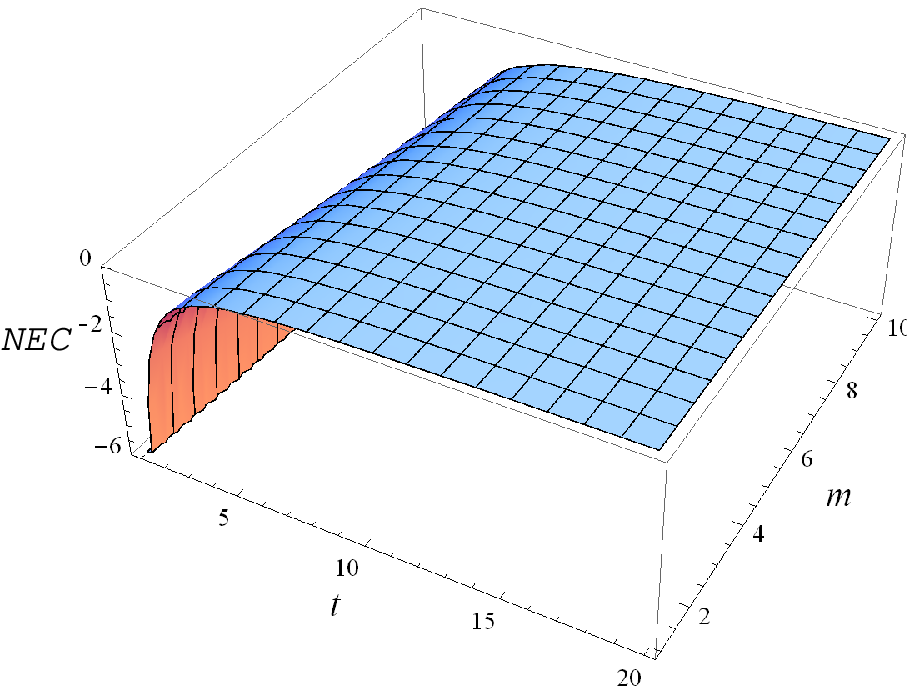, width=.495\linewidth,
height=2.2in} \epsfig{file=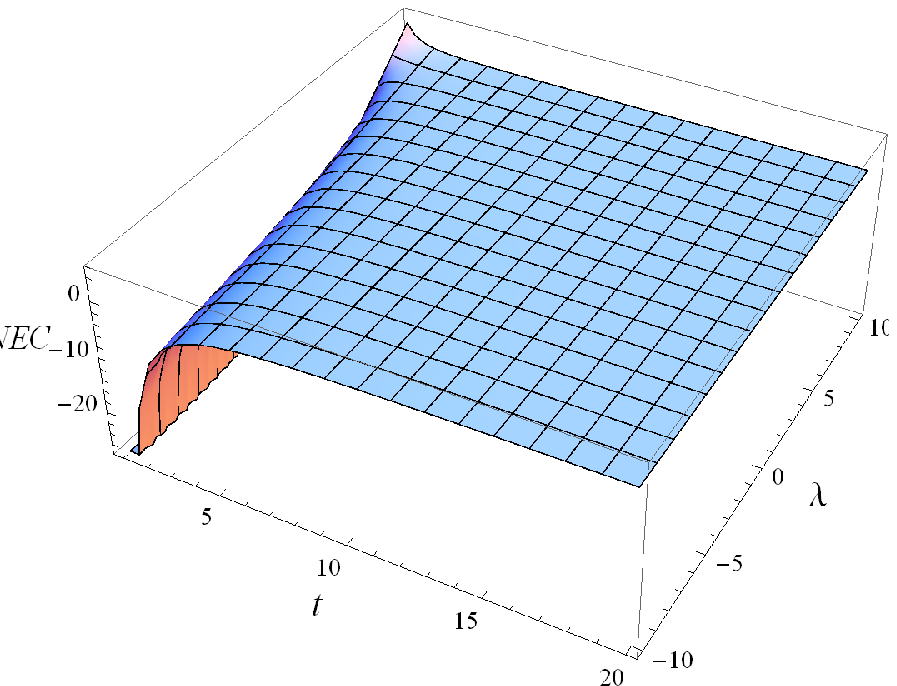, width=.495\linewidth,
height=2.2in} \caption{(Colour online) Evolution of NEC for $B(R)$
in HDE (a) with $\lambda=0.1$ and varying $m$ (b) with $m=10$ and
varying $\lambda$.}
\end{figure}
\begin{figure}
\centering \epsfig{file=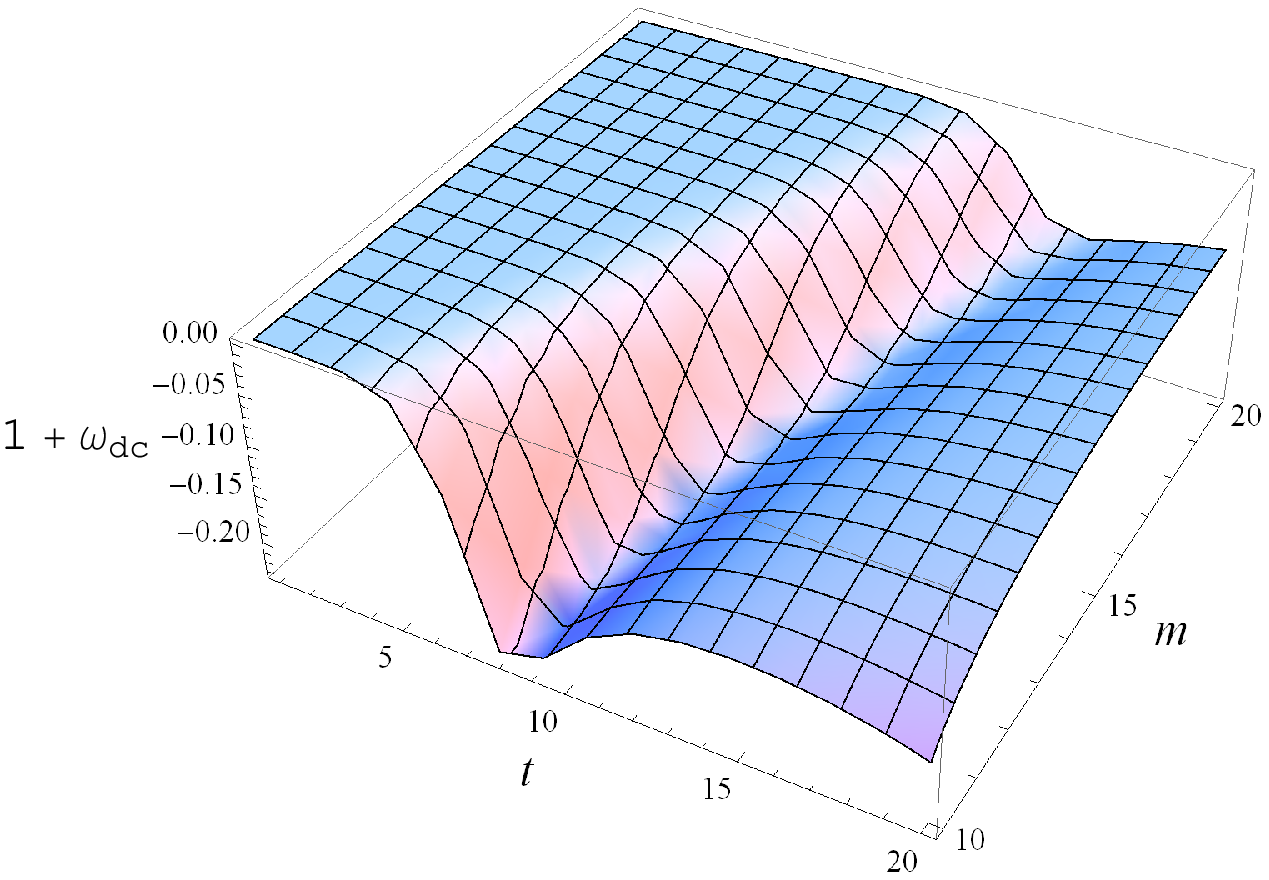, width=.495\linewidth,
height=2.2in} \epsfig{file=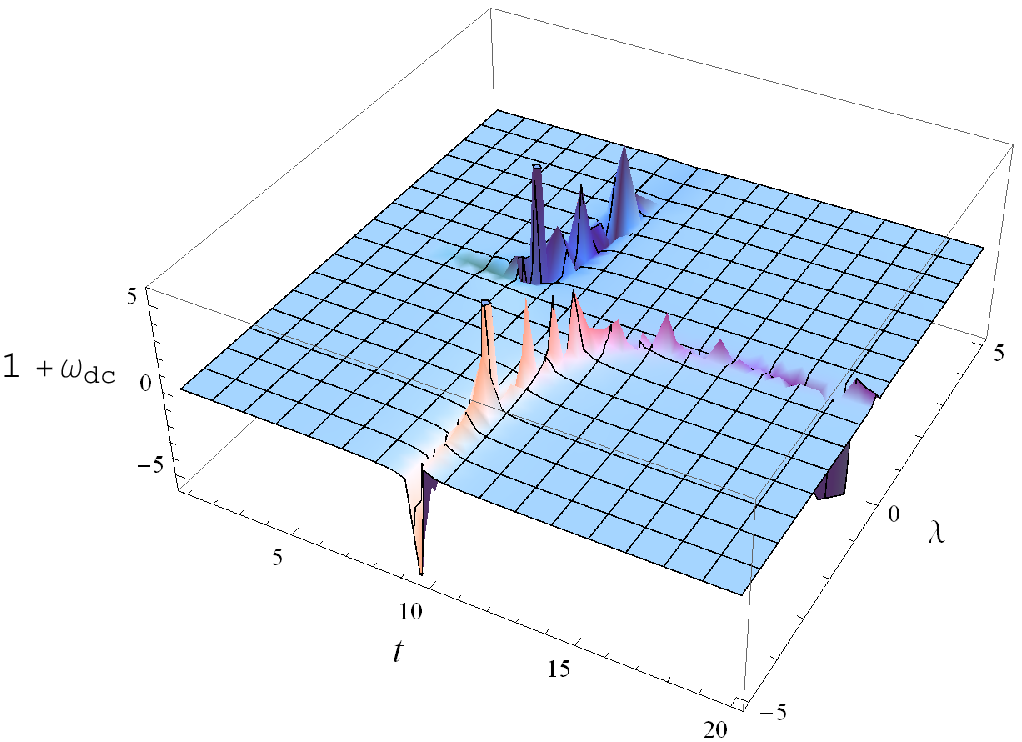, width=.495\linewidth,
height=2.2in} \caption{(Colour online) Evolution of $1+\omega_{dc}$
for $B(R)$ in HDE (a) with $\lambda=0.1$ and varying $m$ (b) with
$m=10$ and varying $\lambda$.}
\end{figure}

We also explore the behavior of NEC for the reconstructed $B(R)$ in
HDE and display the graphs for different values of parameters $m$
and $\lambda$. Figure \textbf{5} shows that NEC is violated
\emph{i.e.}, $\rho_{dc}+p_{dc}<0$ which necessitates
$\omega_{dc}<-1$. To make sure the phantom regime of the DE, we also
plot the evolution of $1+\omega_{dc}$ against $m$ and $\lambda$
shown in Figure \textbf{6}. The plots clearly favors the accelerated
expansion except for particular range of $\lambda$. Thus the
$f(R,T)$ model corresponding to HDE is consistent with present day
observations$^{{1-2})}$.

\section{Reconstruction from New Agegraphic Dark Energy}

In this section, we discuss the reconstruction of $f(R,T)$ gravity
in the setting of NADE. The energy density of NADE is proposed as
$^{{30})}$
\begin{equation}\label{41}
\rho_{\vartheta}=\frac{3n^2M^{2}_p}{\xi^2},
\end{equation}
where the numerical component $3n^2$ is inserted to parameterize
some uncertainties namely, the specific forms of cosmic quantum
fields and the role of curvature of spacetime etc., $\xi$ is the
conformal time in FRW background defined as
\begin{equation*}
\xi=\int{\frac{dt}{a(t)}} =\int{\frac{da}{Ha^{2}}}.
\end{equation*}
Wei and Cai$^{{30})}$ developed the cosmological constraints on NADE
and found that the resolution of coincidence problem may become more
definite in the NADE model with specific value of $n$ nearly unity.
They constrained the NADE by using the observational data of SNeIa,
CMB and LSS and found the best fit parameter (with $1\sigma$
uncertainty) $n=2.76^{+0.111}_{-0.109}$. The new agegraphic DE has
been under consideration in both GR and modified theories
scenario$^{{31})}$. The time derivative of $\rho_{\vartheta}$ is
obtained as
\begin{equation}\label{42}
\dot{\rho}_{\vartheta}=\frac{-2\rho_{\vartheta}H\sqrt{\Omega_{\vartheta}}}{an}.
\end{equation}
Substituting Eq.(\ref{42}) in Eq.(\ref{21}), it follows that
\begin{equation}\label{43}
\omega_{\vartheta}=-1+\frac{2}{3n}\frac{\sqrt{\Omega_{\vartheta}}}{a}.
\end{equation}
We are concerned to demonstrate the possible correspondence between
$f(R,T)$ models and NADE. In the following, we discuss the two cases
individually.
{\begin{itemize}
\item {$R+2A(T)$}
\end{itemize}}
Comparing Eqs.(\ref{43}) and (13), we obtain
\begin{equation}\label{44}
TA_T-\frac{\sqrt{\Omega_{\vartheta}}}{3na-2\sqrt{\Omega_{\vartheta}}}A=0.
\end{equation}
For $a(t)=a_0(t_p-t)^{-m}$, its solution is $A(T)=C_4T^{K_1}$, where
$C_4$ is constant of integration and $K_1=(m+1)/(m-2)$. Now, we
develop initial constraint on $A(T)$ for NADE model and find out the
constant $C_4$. Evaluating Eq.(\ref{44}) at present day and
manipulating with Eq.(\ref{30}), we obtain the following initial
condition on $A(T)$
\begin{equation}\label{45}
A(T_0)=3H_0^2{\Omega}_{\vartheta0}\left(1+\frac{2\sqrt{\Omega_{\vartheta0}}}
{3na_0-2\sqrt{\Omega_{\vartheta0}}}\right)^{-1}.
\end{equation}
Making use of Eq.(\ref{45}) and relation $A(T)=C_4T^{K_1}$, the
$f(R,T)$ model is constructed as
\begin{equation}\label{46}
f(R,T)=R+6H_0^2{\Omega}_{\vartheta0}T_0^{-K_1}\left(1+\frac{2\sqrt{\Omega_{\vartheta0}}}
{3na_0-2\sqrt{\Omega_{\vartheta0}}}\right)^{-1}T^{K_1}.
\end{equation}
\begin{figure}
\centering \epsfig{file=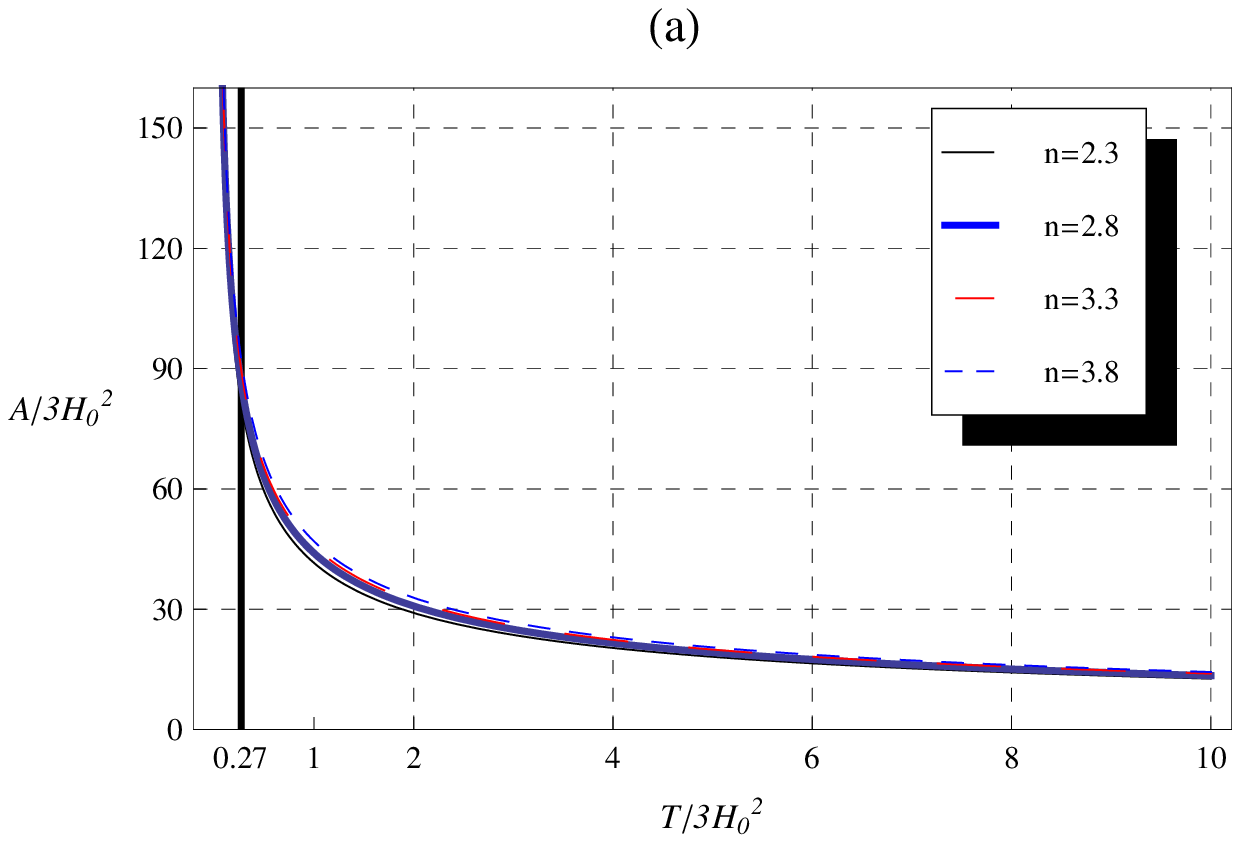, width=.495\linewidth,
height=2.2in} \epsfig{file=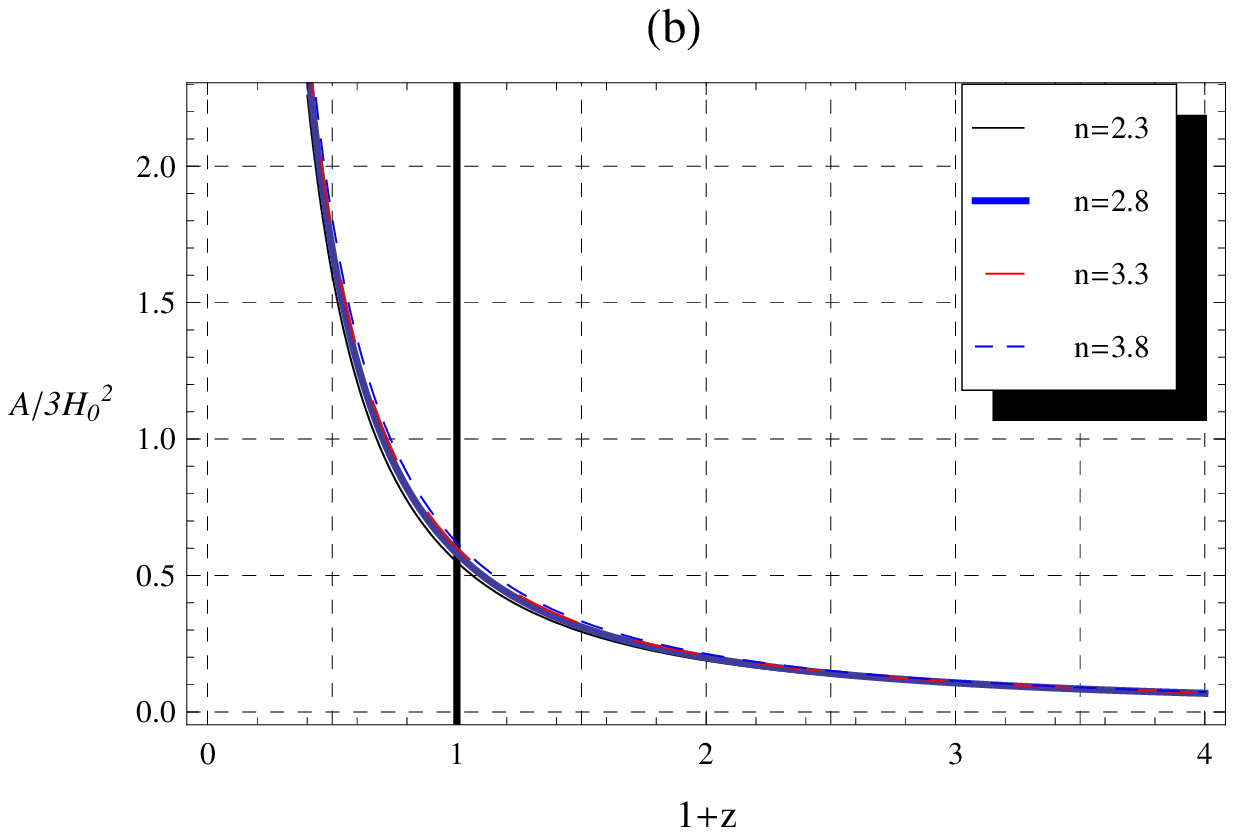, width=.495\linewidth,
height=2.2in} \caption{(Colour online) Evolution of $A(T)$ in NADE
(a) versus $T$ and (b) versus $z$ for different values of $n$. Black
thick line represents the current value of $T=T_0$.}
\end{figure}
\begin{figure}
\centering \epsfig{file=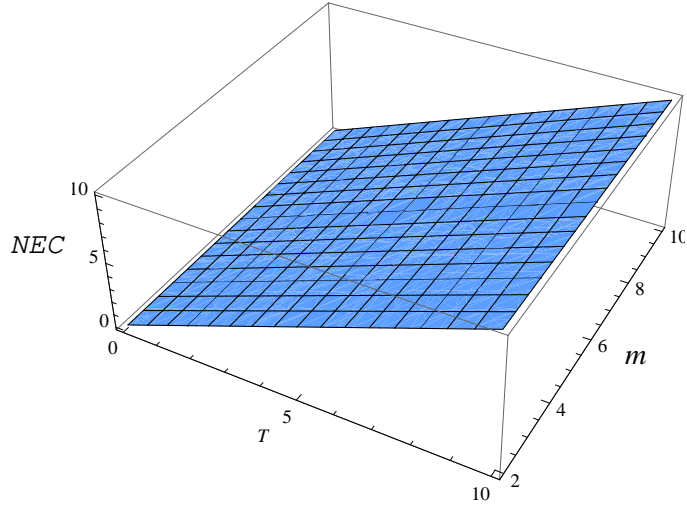}\caption{(Colour online) Evolution
of NEC in NADE versus $T$ and $m$.}
\end{figure}

In case of NADE, we set $n=2.3, 2.8, 3.3, 3.8$ and plot $A(T)$ in
terms of $T$ and redshift as shown in Figure \textbf{7}. One can see
that the difference in evolutionary curves of $A(T)$ depending on
the value of $n$ is not obvious in both graphs. These plots
represent the future era where $A(T)$ is increasing rapidly. The EoS
of DE components for the $A(T)$ model (\ref{46}) is found as
$\omega_{dc}=-1+\frac{2(m+1)}{3m}$ which represents the quintessence
era of DE. We also plot the NEC for this $A(T)$ model by varying the
values of parameter $m$ shown in Figure \textbf{8}. The NEC is found
to be satisfied \emph{i.e.}, $\rho+p>0$ which confirms the regime
with $\omega_{dc}>-1$. Hence, the reconstructed $A(T)$ for NADE
represents the quintessence era of the universe. {\begin{itemize}
\item {$B(R)+{\lambda}T$}
\end{itemize}}
The conformal time $\xi$ of FRW universe can be represented in terms
of Ricci scalar $R$ as
\begin{equation*}
\xi=\frac{1}{a_0(m+1)}\left[\frac{6m(2m+1)}{R}\right]^{\frac{m+1}{2}}.
\end{equation*}
Likewise $\rho_M$ and $(1+\omega_{\vartheta})\rho_{\vartheta}$ for
NADE are determined as
\begin{eqnarray}\label{47}
\rho_M=\frac{3M^2_p[n^2a^2_0(m+1)^2(-1)^mR^m-m^2(6m(2m+1))^m]}
{[6m(2m+1)]^{m+1}}R,\\\label{48}
(1+\omega_{\vartheta})\rho_\vartheta=\frac{2n^2a_0^2M^{2}_p(m+1)^3(-1)^{m+1}}
{m[6m(2m+1)]^{m+1}}R^{m+1}.
\end{eqnarray}
Solving the differential equation (\ref{17}) for NADE, it follows
that
\begin{equation}\label{49}
B(R)=\mu_{-}R^{\jmath_{-}}C_5+\mu_{+}R^{\jmath_{+}}C_6+\chi{R}^m+\gamma{R}+C_7,
\end{equation}
where $\gamma$ and $\chi$ are given by
\begin{eqnarray*}
\gamma&=&\frac{1}{2m}[2m-3\lambda{M_p^2}m^2],\\\nonumber
\chi&=&\frac{n^2a_0^2(m+1)(-1)^{m+1}[3\lambda{M^2_p}m+2(m+1)]}{2(m^3+2m^2)[6m(2m+1)]^{m}}.
\end{eqnarray*}
Here, constants $C_5,~C_6$ and $C_7$ can be determined from the
initial conditions (\ref{37}) and (\ref{38}). The resulting NADE
model of the Lagrangian $B(R)+{\lambda}T$ is
\begin{equation}\label{50}
f(R,T)=\chi{R}^m+C_{+}R^{\jmath_{+}}+C_{-}R^{\jmath_{-}}+\gamma{R}+\delta+{\lambda}T,
\end{equation}
where
\begin{eqnarray*}
C_{+}&=&\frac{(\gamma-1)(\jmath_{-}-1)R_0+\chi{m}(\jmath_{-}-m)\chi{R}_0^m}
{\jmath_{+}(\jmath_{+}-\jmath_{-})R_0^{\jmath_{+}}},
\\\nonumber
C_{-}&=&\frac{(\gamma-1)(\jmath_{+}-1)R_0+\chi{m}(\jmath_{+}-m)\chi{R}_0^m}
{\jmath_{-}(\jmath_{-}-\jmath_{+})R_0^{\jmath_{-}}},
\\\nonumber
\delta&=&\beta+(1-\gamma)R_0-\chi{R_0}^m+\frac{1}{\jmath_+\jmath_-}
\left[(\gamma-1)(\jmath_++\jmath_--1)R_0\right.\nonumber\\&+&\left.m\chi(\jmath_++\eta_--1)R_0^m\right].
\end{eqnarray*}
\begin{figure}
\centering \epsfig{file=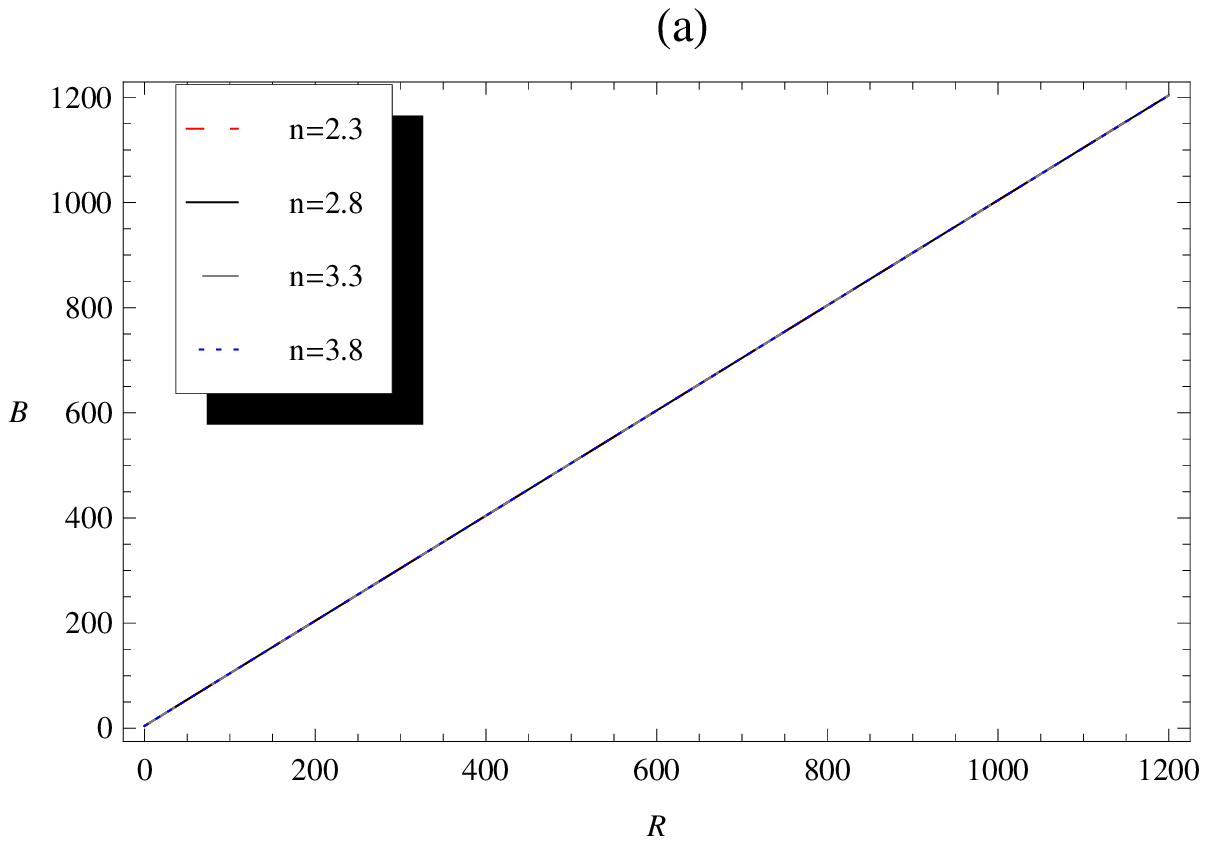, width=.495\linewidth,
height=2.2in} \epsfig{file=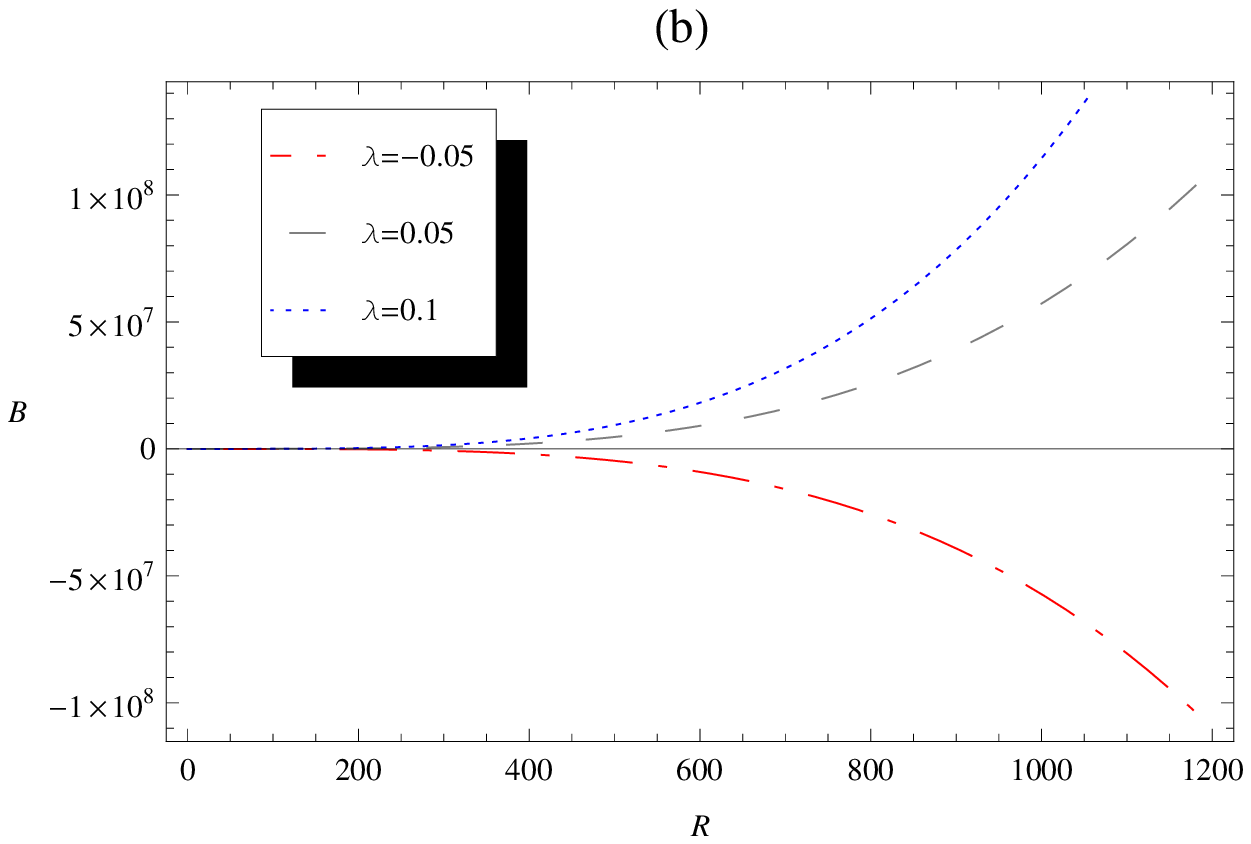, width=.495\linewidth,
height=2.2in} \caption{(Colour online) Evolution of $B(R)$ versus
$R$ in NADE for (a) different values of $n$ with $\lambda=0$ and (b)
different values of $\lambda$ with $n=2.8$.}
\end{figure}

For the NADE, the function $B(R)$ is plotted against $R$ for
different values of parameter $n$ and $\lambda$ as shown in Figure
\textbf{9}. In Figure \textbf{9(a)}, we fix $\lambda=0$ (corresponds
to $f(R)$ gravity) and represent the behavior of $B(R)$ for
different values of $n$. It shows that the curves for reconstructed
$B(R)$ in NADE are same. If one introduces the coupling parameter
$\lambda$ with $n=2.8$, the variation in results is evident from
Figure \textbf{9(b)}. We also plot these results in $B-z$ plane and
represent the future evolution of $B(R)$ as shown in Figure
\textbf{10}.
\begin{figure}
\centering \epsfig{file=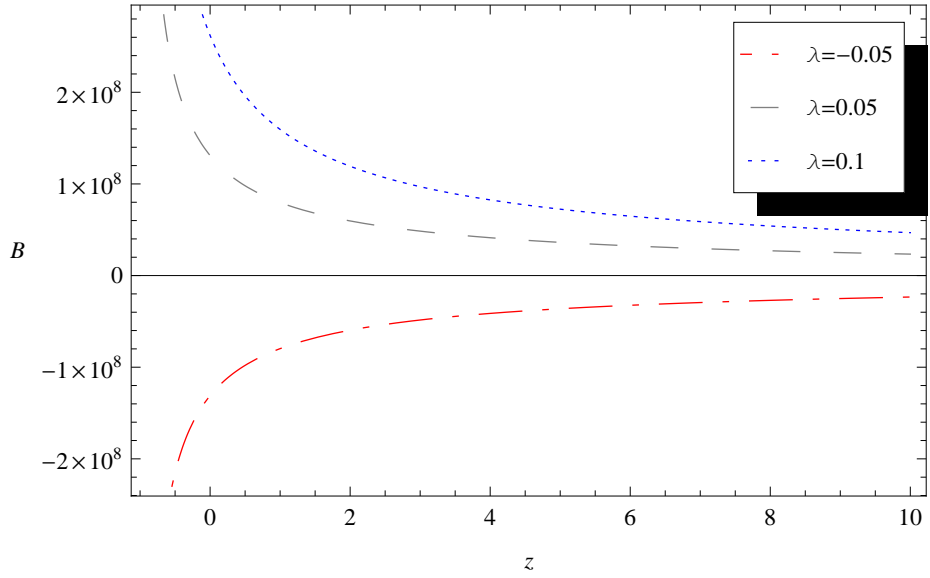}\caption{(Colour online) Evolution
of $B(R)$ versus $z$ in NADE.}
\end{figure}

Now we check the validity of NEC for the $B(R)$ in NADE shown in
Figure \textbf{11}. It is clear that NEC is satisfied \emph{i.e.},
$\rho_{dc}+p_{dc}>0$ except for the negative values of coupling
parameter $\lambda$. Consequently, these models should imply
$\omega_{dc}>-1$, the quintessence EoS parameter. We show the
evolution of $1+\omega_{dc}$ for different values of parameters $m$
and $\lambda$. The plots in Figure \textbf{12} make it more definite
that the reconstructed function $B(R)$ favors the quintessence
regime of the universe.
\begin{figure}
\centering \epsfig{file=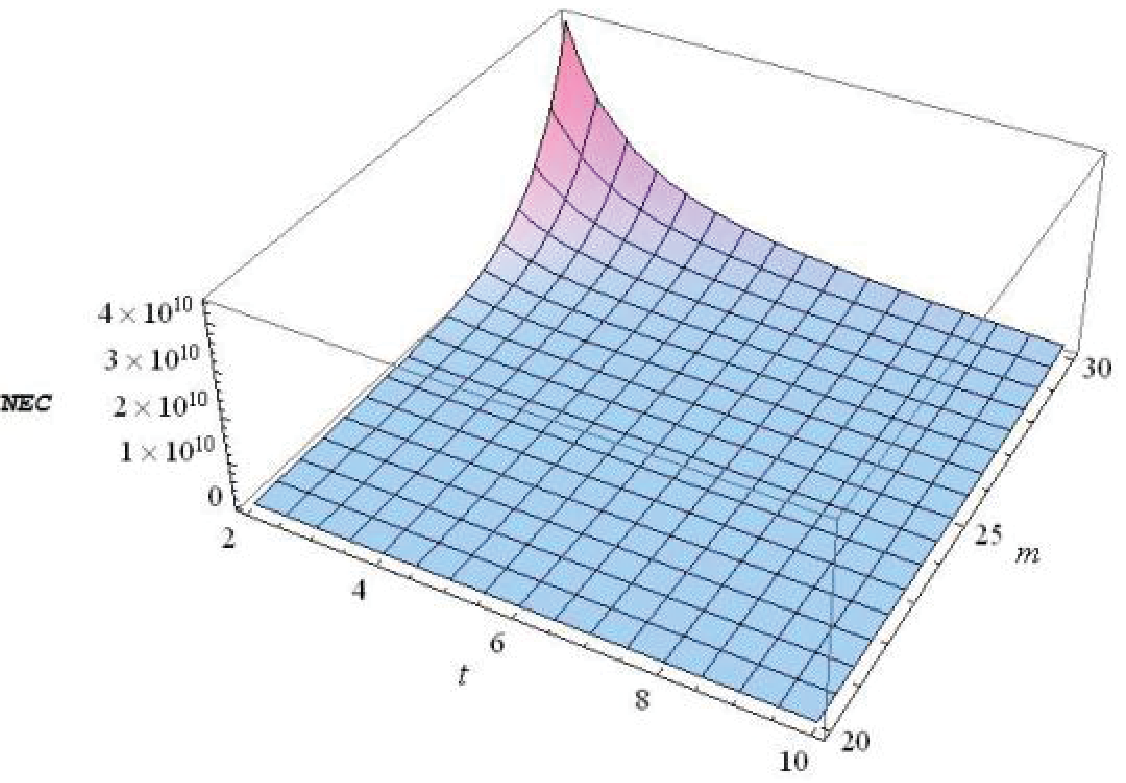, width=.495\linewidth,
height=2.2in} \epsfig{file=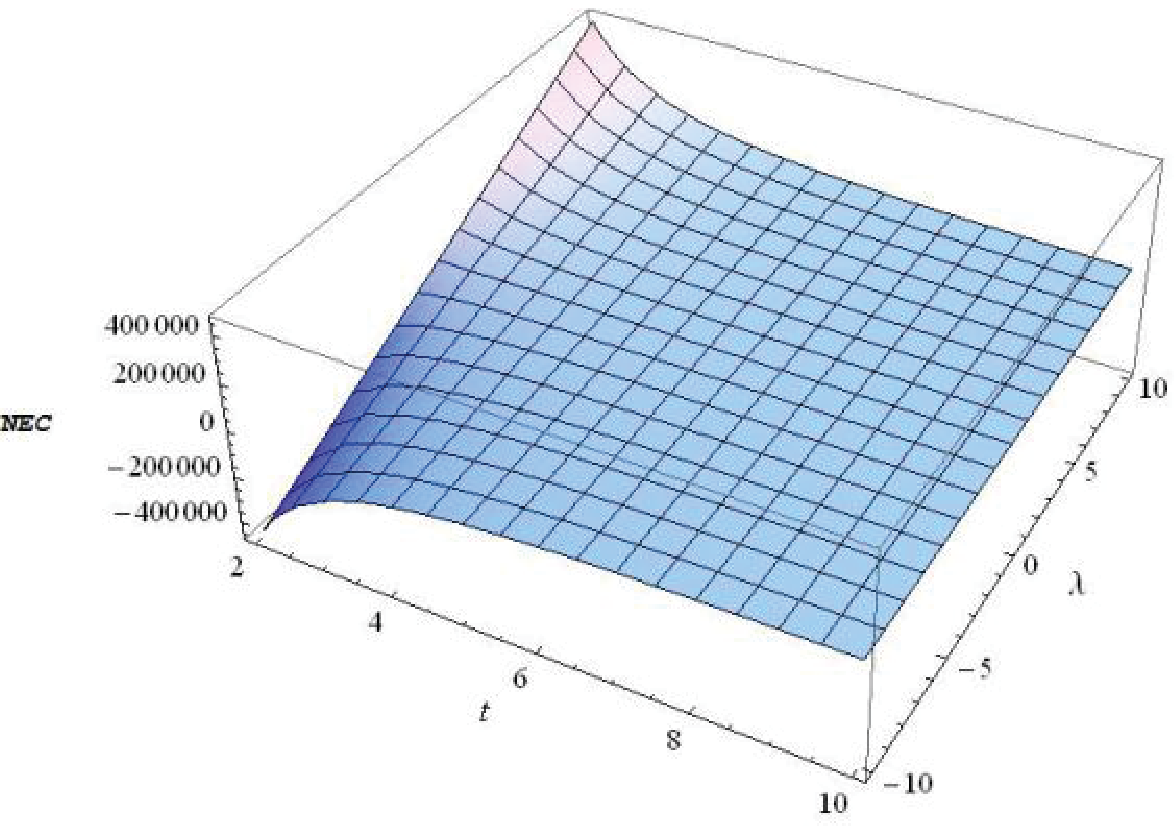, width=.495\linewidth,
height=2.2in} \caption{(Colour online) Evolution of NEC for $B(R)$
in NADE (a) with $\lambda=0.1$ and varying $m$ (b) with $m=10$ and
varying $\lambda$.}
\end{figure}
\begin{figure}
\centering \epsfig{file=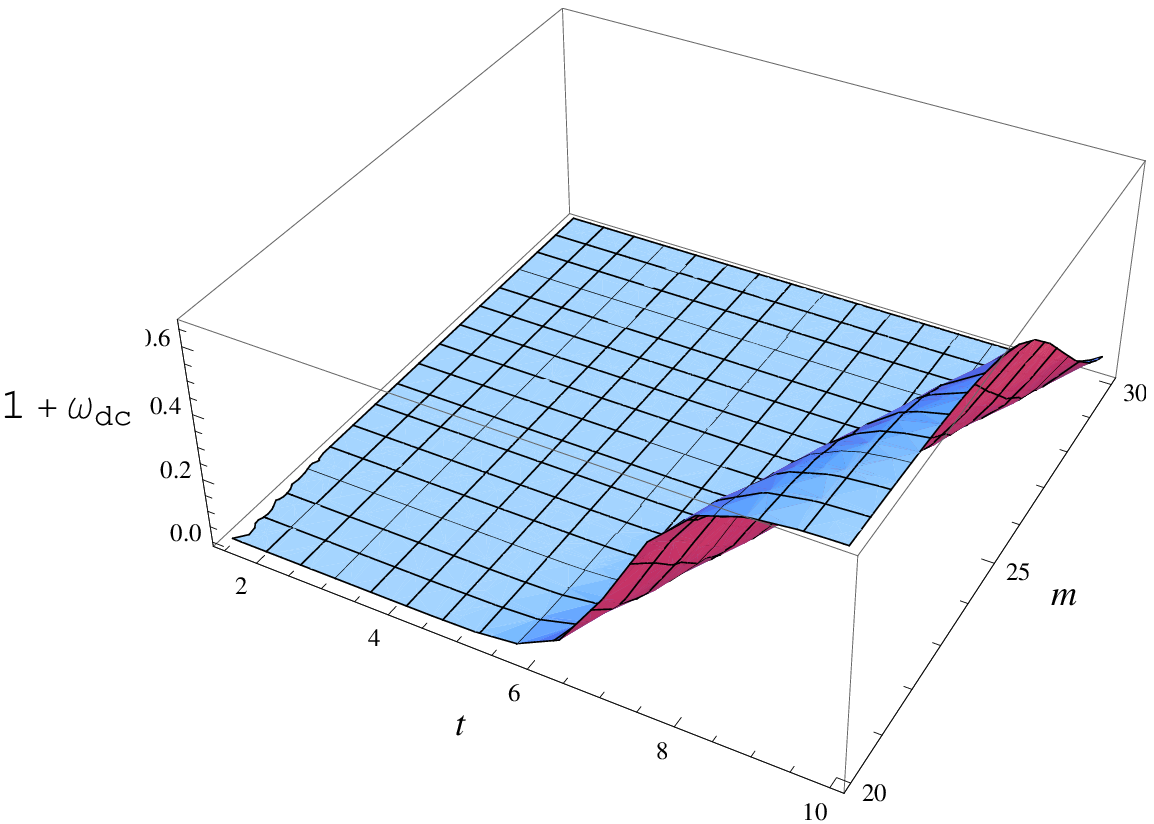, width=.498\linewidth,
height=2.3in}\epsfig{file=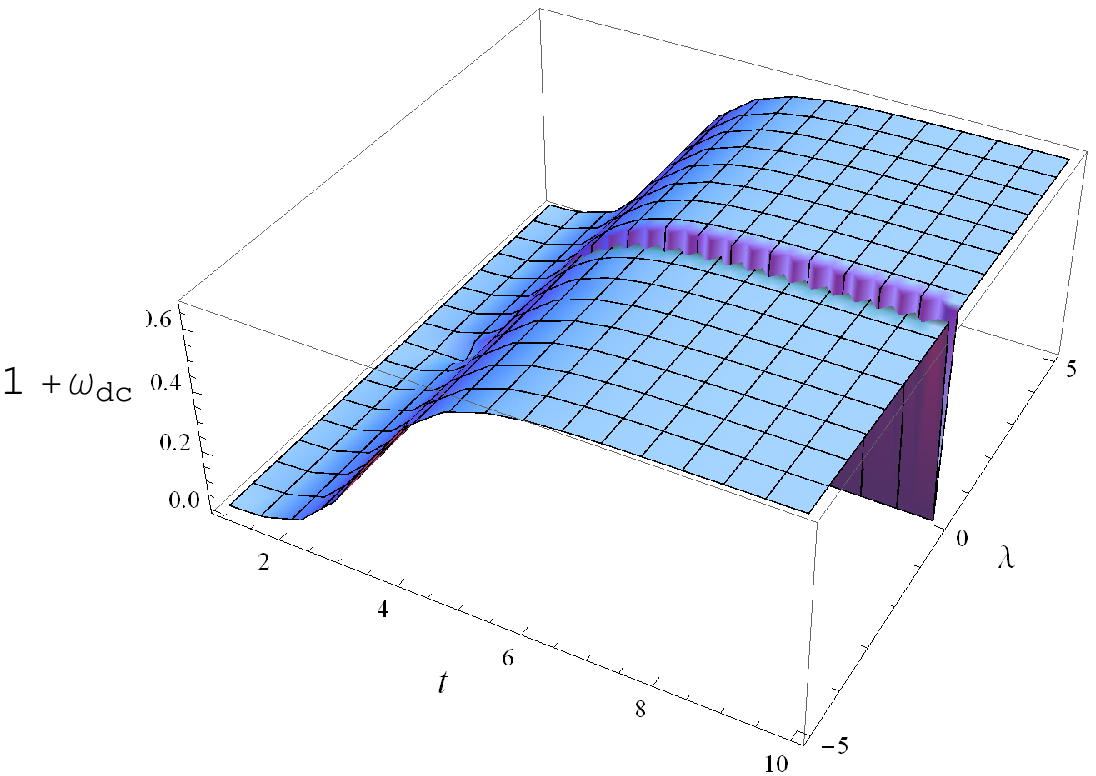, width=.498\linewidth,
height=2.3in} \caption{(Colour online) Evolution of $1+\omega_{dc}$
for $B(R)$ in NADE (a) with $\lambda=0.1$ and varying $m$ (b) with
$m=10$ and varying $\lambda$.}
\end{figure}

\section{Generalized Second Law of Thermodynamics}

Here, we discuss the validity of GSLT in this modified gravity on
the future event horizon. The GSLT states that entropy of a black
hole horizon summed to the entropy of matter and fluids inside the
horizon is non-decreasing with time. The validity of GSLT has been
discussed in the setting of modified theories of gravity
$^{{32-34})}$. In$^{15)}$, a non-equilibrium picture of
thermodynamics is discussed on the apparent horizon of FRW spacetime
in $f(R,T)$ gravity. It is remarked that usual laws of
thermodynamics do not hold in this modified theory and additional
entropy production term $\hat{S}_\jmath$ is required. We consider a
flat FRW universe consisting of ordinary matter plus the DE
component. The modified first law of thermodynamics is stated
as$^{15)}$
\begin{equation}\label{53}
T_hd\hat{S}_{in}=Vd\rho_{EFF}+(\rho_{EFF}+p_{EFF})dV-T_hd\hat{S}_\jmath,
\end{equation}
where $T_{h}$ and $\hat{S}_{in}$ represent temperature and entropy
of entire contents within the horizon. We have to show that
\begin{equation}\label{54}
\dot{\hat{S}}=\dot{\hat{S}}_h+\dot{\hat{S}}_{in}+\dot{\hat{S}}_\jmath\geqslant0,
\end{equation}
where $\hat{S}_h$ is the horizon entropy. For
$V=4\pi{R}_{\hat{E}}^3/3$, Eq.(\ref{53}) yields
\begin{equation}\label{55}
T_h\dot{\hat{S}}_{in}=\frac{4}{3}\pi{R}_{\hat{E}}^3\dot{\rho}_{EFF}
+4\pi(\rho_{EFF}+p_{EFF})\dot{R}_{\hat{E}}\dot{R}_{\hat{E}}-T_h\dot{\hat{S}}_\jmath,
\end{equation}
We assume that temperature $T_h$ is proportional to Gibbson-Hawking
temperature$^{{33, 35})}$
\begin{equation}\label{56}
T_h=\frac{lH}{2\pi},
\end{equation}
where $l$ is a real constant. In the following, we study GSLT for
two forms of $f(R,T)$ function.
{\begin{itemize}
\item {$f(R,T)=R+2A(T)$}
\end{itemize}}
In GR, the Bekenstein-Hawking entropy is given by the relation
$\hat{S}_h=\hat{A}/4G$, where $\hat{A}=4{\pi}R_{\hat{E}}^2$
represents the area of the event horizon$^{{36})}$. It was proposed
that the horizon entropy is associated with Noether charge in the
context of modified gravity theories$^{{37})}$. Brustein et al.
$^{{38})}$ interpreted that Wald entropy is equivalent to one-fourth
of the horizon area with gravitational coupling being the effective
one. Hence, the entropy in this modified gravity is defined as
$^{{15})}$
\begin{equation}\label{57}
{\hat{S}}_h=\frac{\hat{A}}{4G_{EFF}}, \quad G_{EFF}=G+2A_T/8\pi,
\end{equation}
its time rate is
\begin{equation}\label{58}
\dot{\hat{S}}_h=\left(2\pi{R}_{\hat{E}}\dot{R}_{\hat{E}}
+\pi{R}_{\hat{E}}^2\frac{d}{dt}\right)\frac{1}{G_{EFF}}.
\end{equation}
Using the FRW equations for this $f(R,T)$ model, Eq.(\ref{55}) leads
to
\begin{equation}\label{59}
\dot{\hat{S}}_{in}+\dot{\hat{S}}_\jmath=\frac{2\pi{R}_{\hat{E}}^2}{lH}\left(\dot{H}
+\frac{H^2R_{\hat{E}}}{2}\frac{d}{dt}\right)\frac{1}{G_{EFF}},
\end{equation}
Thus, the total entropy for GSLT becomes
\begin{equation}\label{60}
\dot{\hat{S}}=\frac{\pi{R}_{\hat{E}}^2}{G_{EFF}}\left[2\left(\frac{\dot{R}_{\hat{E}}}{{R}_{\hat{E}}}
+\frac{\dot{H}}{lH}\right)-\left(1+\frac{HR_{\hat{E}}}{l}\right)
\frac{\dot{G}_{EFF}}{G_{EFF}}\right]\geqslant0,
\end{equation}
or equivalently
\begin{equation}\label{61}
\dot{\hat{S}}=\frac{2\pi{R}_{\hat{E}}^2}{G_{EFF}}\left[\frac{d}{dt}[ln(R_{\hat{E}}H^{1/l})]
+\ln
e^{(1+\frac{HR_{\hat{E}}}{l})}\frac{d}{dt}[\ln\frac{1}{\sqrt{G_{EFF}}}]\right]\geqslant0.
\end{equation}

In GR, the above condition reduces to
$(R_{\hat{E}}H^{1/l})\geqslant0$. The effective gravitational
coupling constant for this $f(R,T)$ model needs to be positive so
that $A_T>0$. To illustrate our result, let us consider the $f(R,T)$
model given by Eq.(\ref{33}). In this model, $H=\frac{m}{(t_p-t)}$,
$\dot{H}=\frac{m}{t_p-t}$ and $R_{\hat{E}}=\frac{(t_p-t)}{(m+1)}$.
By the direct replacement of these results, we obtain that GSLT is
valid if $l\leqslant1$, $A_T>0$ and $\dot{A}_T\leqslant0$. For
$A(T)=3H_0^2{\Omega}_{\vartheta0}T_0^{-K}\left(1+2\frac{e-\sqrt{\Omega_{\vartheta0}}}
{e+2\sqrt{\Omega_{\vartheta0}}}\right)^{-1}T^K$, the condition
$A_T>0$ holds if
$3H_0^2{\Omega}_{\vartheta0}T_0^{-K}\left(1+2\frac{e-\sqrt{\Omega_{\vartheta0}}}
{e+2\sqrt{\Omega_{\vartheta0}}}\right)^{-1}<(3m+2){T}^{1-K}$ and
$\dot{A_T}<0$, since
$\dot{A_T}=\dot{T}K(K-1)3H_0^2{\Omega}_{\vartheta0}T_0^{-K}\left(1+2\frac{e-\sqrt{\Omega_{\vartheta0}}}
{e+2\sqrt{\Omega_{\vartheta0}}}\right)^{-1}T^{K-2}$ with $\dot{T}=\dot\rho<0$.\\
{\begin{itemize}
\item {$f(R,T)=B(R)+{\lambda}T$}
\end{itemize}}
For this specific model, the Wald entropy is defined as$^{{15})}$
\begin{equation}\label{62}
\hat{S}_h=\frac{\hat{A}B_R}{4\tilde{G}}, \quad
\tilde{G}=G+\lambda/8\pi,
\end{equation}
whose time derivative gives
\begin{equation}\label{63}
\dot{\hat{S}}_h=\left(2\pi{R}_{\hat{E}}\dot{R}_{\hat{E}}B_R+\pi{R}_{\hat{E}}^2\dot{B_R}\right)\frac{1}{\tilde{G}}.
\end{equation}
Following the above procedure, the GSLT leads to
\begin{equation}\label{64}
\frac{2\pi{R}_{\hat{E}}^2B_R}{\tilde{G}}\left[\frac{d}{dt}[\ln(R_{\hat{E}}H^{1/l})]
+\ln
e^{(1+\frac{HR_{\hat{E}}}{l})}\frac{d}{dt}[\ln\sqrt{B_R}]\right]\geqslant0.
\end{equation}
For the particular choice of scale factor $a(t)=a_0(t_p-t)^{-m}$
with $R=\frac{-6m(2m+1)}{(t_p-t)^2}$, we consider the
$f(R,T)=B(R)+{\lambda}T$ model (\ref{40}) corresponding to HDE. The
GSLT would be valid if $l\leqslant1$ and scalar curvature lies in
the range
$\frac{-\gamma{R}^{1-\jmath_-}-C_{-}\jmath_{-}}{C_+\jmath_+}<R^{\jmath_+-\jmath_-}
<\frac{C_-\jmath_-(1-\jmath_-)}{C_+\jmath_+(\jmath_+-1)}$.

\section{Conclusions}

The $f(R,T)$ theory can be reckoned as a useful candidate of
\emph{dark energy components} which may help to understand the
accelerated expansion of the universe. In such theory, cosmic
acceleration may appear as an outcome of unified contribution from
geometrical and matter components. We have discussed the
cosmological reconstruction of $f(R,T)$ theory in the light of
holographic and new agegraphic DE models. There are various models
of $f(R,T)$ Lagrangian$^{{14})}$ but we have concentrated on
$f(R,T)=f_1(R)+f_2(T)$ with particular functions $f_1$ and $f_2$.
The model $f(R,T)=R+2A(T)$ matches the usual Einstein action plus
time dependent cosmological constant which is presented as function
of trace of the energy-momentum tensor. One can see that if the
contribution of curvature matter coupling is null, i.e., $A(T)=0$
then the model reduces to GR which represents the matter dominated
universe. The second model $f(R,T)=B(R)+{\lambda}T$ appears as
matter corrected $f(R)$ type gravity.

We have formulated the field equations for each model in flat FRW
background and obtained the evolution equation for the respective
unknown functions. The HDE and NADE models are proposed as an
equivalent description to DE components originating from the stated
modified theory. Some analytical solutions have been obtained by
applying the initial conditions on respective functions.
Accordingly, one can determine the explicit $f(R,T)$ functions
corresponding to HDE and NADE.

For HDE dominated universe, \emph{i.e.},
$\Omega_{\vartheta}\thicksim1$; if $e>1$ then expansion is in
quintessence regime and Eq.({\ref{25}}) implies that
$A(T)\propto{T}^\alpha,~\alpha>0,~e=1$ leading to the de Sitter
universe with $A(T)\propto{constant}$. When $e<1$, phantom evolution
of the universe is on cards with $A(T)\propto{T}^\alpha,~\alpha<0$.
The reconstructed $A(T)$ model satisfies the EoS parameter
$\omega_{dc}<-1$ which is evident from Figure \textbf{2}. For the
model $f(R,T)=B(R)+{\lambda}T$, we discuss the evolution of $B(R)$
and explore the behavior of NEC and $1+\omega_{dc}$. The NEC is
found to be violated which results in $\omega_{dc}<-1$ as depicted
in Figures \textbf{5} and \textbf{6}. Thus the $f(R,T)$ models
reconstructed for HDE represent the phantom era of DE which is
consistent with the recent observations$^{{1, 2})}$.

In case of NADE having $\Omega_{\vartheta}\thicksim1$, the EoS
parameter
$\omega_{\vartheta}=-1+\frac{2}{3n}\frac{\sqrt{\Omega_{\vartheta}}}{a}$
can be less than $-1$ if $n<0$ but from observational point of view
$n=2.76^{+0.111}_{-0.109}$$^{{30})}$ which permits the quintessence
era and the corresponding $A(T)$ model is of the form
$A(T)\propto{T}^{\frac{1}{3na-1}}$. The EoS parameter corresponding
to $A(T)$ represents the quintessence regime of DE which constitutes
the relation $\rho_{dc}+p_{dc}>0$ as depicted in Figure \textbf{8}.
The evolution of the function $B(R)$ corresponding to NADE $f(R,T)$
model is discussed in Figures \textbf{9-12}. These plots show the
influence of coupling parameter $\lambda$ on the evolutionary regime
of the universe. We find that function $B(R)$ for the NADE favors
the quintessence era of the DE.

The EoS parameter $\omega_{dc}$ for the above $f(R,T)$ models is in
agreement with the observational data of WMAP5$^{{39})}$. Hence, we
can suggest that these reconstructed models of $f(R,T)$ gravity are
consistent with the evolution of HDE and NADE in general relativity.
The polynomial functions (\ref{40}) and (\ref{50}) represent more
general $f(R,T)$ models of the type $B(R)+{\lambda}T$. If one puts
$\lambda=0$ then the respective models in $f(R)$ gravity can be
reproduced. Though $f(R)$ theory has been reconstructed for HDE and
NADE but these functions appear to be more general. We have also
assured the validity of GSLT on the future event horizon of FRW
universe. The HDE $f(R,T)$ models are employed to establish the
constraints which validate the GSLT in this modified gravity.

\vspace{.5cm}

{\bf Acknowledgment}

\vspace{0.25cm}

We would like to thank the Higher Education Commission, Islamabad,
Pakistan for its financial support through the {\it Indigenous Ph.D.
5000 Fellowship Program Batch-VII}. The authors are grateful to the
Physical Society of Japan for Financial Support in publication\\
1) S. Perlmutter, S. Gabi, G. Goldhaber, A. Goobar, D. E. Groom, I.
M. Hook, A. G. Kim, M. Y. Kim, G. C. Lee, R. Pain, C. R.
Pennypacker, I. A. Small, R. S. Ellis, R. G. McMahon, B. J. Boyle,
P. S. Bunclark, D. Carter, M. J. Irwin, K. Glazebrook, H. J. M.
Newberg, A. V. Filippenko, T. Matheson, M. Dopita and W. C. Couch:
Astrophys. J. \textbf{483} (1997) 565; A. G. Riess, L. G. Strolger,
J. Tonry, Z. Tsvetanov, S. Casertano, H. C. Ferguson, B. Mobasher,
P. Challis, N. Panagia, A. V. Filippenko, W. Li, R. Chornock, R. P.
Kirshner, B. Leibundgut, M. Dickinson, A. Koekemoer, N. A. Grogin
and M. Giavalisco : Astrophys. J. \textbf{607} (2004) 665.\\
2) C. L. Bennett, M. Halpern, G. Hinshaw, N. Jarosik, A. Kogut, M.
Limon, S. S. Meyer, L. Page, D. N. Spergel, G. S. Tucker, E.
Wollack, E. L. Wright, C. Barnes, M. R. Greason, R. S. Hill, E.
Komatsu, M. R. Nolta, N. Odegard, H. V. Peiris, L. Verde and J. L.
Weiland: Astrophys. J. Suppl. \textbf{148} (2003) 1; D. N. Spergel,
R. Bean, O. Doré, M. R. Nolta, C. L. Bennett, J. Dunkley, G.
Hinshaw, N. Jarosik, E. Komatsu, L. Page, H. V. Peiris, L. Verde, M.
Halpern, R. S. Hill, A. Kogut, M. Limon, S. S. Meyer, N. Odegard, G.
S. Tucker, J. L. Weiland, E. Wollack and E. L. Wright: Astrophys. J.
Suppl. \textbf{170} (2007) 377.\\
3) E. Hawkins, S. Maddox, S. Cole, O. Lahav, D. S. Madgwick, P.
Norberg, J. A. Peacock, I. K. Baldry, C. M. Baugh, J.
Bland-Hawthorn, T. Bridges, R. Cannon, M. Colless, C. Collins, W.
Couch, G. Dalton, R. D. Propris, S. P. Driver, S. P., G. Efstathiou,
R. S. Ellis, C. S. Frenk, K. Glazebrook, C. Jackson, B. Jones, I.
Lewis, S. Lumsden, W. Percival, B. A. Peterson, W. Sutherland and K.
Taylor: Mon. Not. Roy. Astron. Soc. \textbf{346} (2003) 78; M.
Tegmark, M. A. Strauss, M. R. Blanton, K. Abazajian, S. Dodelson, H.
Sandvik, X. Wang, D. H. Weinberg, I. Zehavi, N. A. Bahcall, F.
Hoyle, D. Schlegel, R. Scoccimarro, M. S. Vogeley, A. Berlind, T.
Budavari, A. Connolly, D. J. Eisenstein, D. Finkbeiner, J. A.
Frieman, J. E. Gunn, L. Hui, B. Jain, D. Johnston, S. Kent, H. Lin,
R. Nakajima, R. C. Nichol, J. P. Ostriker, A. Pope, R. Scranton, U.
Seljak, R. K. Sheth, A. Stebbins, A. S. Szalay, I. Szapudi, Y. Xu,
J. Annis, J. Brinkmann, S. Burles, F. J. Castander, I. Csabai, J.
Loveday, M. Doi, M. Fukugita, B. Gillespie, G. Hennessy, D. W. Hogg,
Z. E. Ivezic´, G. R. Knapp, D. Q. Lamb, B. C. Lee, R. H. Lupton, T.
A. McKay, P. Kunszt, J. A. Munn, L. Connell, J. Peoples, J. R. Pier,
M. Richmond, C. Rockosi, D. P. Schneider, C. Stoughton, D. L.
Tucker, D. E. V. Berk, B. Yanny and D. G. York: Phys. Rev. D
\textbf{69} (2004) 103501.\\
4) D. J. Eisentein,  I. Zehavi, D. W. Hogg, R. Scoccimarro, M. R.
Blanton, R. C. Nichol, R. Scranton, Hee-Jong Seo, M. Tegmark, Z.
Zheng, S. F. Anderson, J. Annis, N. Bahcall, J. Brinkmann, S.
Burles, F. J. Castander, A. Connolly, I. Csabai, M. Doi, M.
Fukugita, J. A. Frieman, K. Glazebrook, J. E. Gunn, J. S. Hendry, G.
Hennessy, Z. Ivezic', S. Kent, G. R. Knapp, H. Lin, Yeong-Shang Loh,
R. H. Lupton, B. Margon, T. A. McKay, A. Meiksin, J. A. Munn, A.
Pope, M. W. Richmond, D. Schlegel, D. P. Schneider, K. Shimasaku, C.
Stoughton, M. A. Strauss, M. SubbaRao, A. S. Szalay, I. Szapudi, D.
L. Tucker, B. Yanny, and D. G. York: Astrophys. J. \textbf{633}
(2005) 560.\\
5) B. Jain and A. Taylor: Phys. Rev. Lett. \textbf{91} (2003)
141302.\\
6) V. Sahni: Lect. Notes Phys. \textbf{653} (2004) 141; M. Sharif
and M. Zubair: Int. J. Mod. Phys. D \textbf{19} (2010) 1957; M. Li,
X.-D. Li, S. Wang and Y. Wang: Commun. Theor. Phys. \textbf{56}
(2011) 525; K. Bamba, S. Capozziello, S. Nojiri and S. D. Odintsov:
Astrophys. Space Sci. \textbf{342} (2012) 155.\\
7) S. Weinberg: Rev. Mod. Phys. \textbf{61} (1989) 1; P. J. E.
Peebles
and B. Ratra: Rev. Mod. Phys. \textbf{75} (2003) 559.\\
8) L. Susskind: J. Math. Phys. \textbf{36} (1995) 6377.\\
9) A. G. Cohen, D. B. Kaplan and A. E. Nelson: Phys. Rev. Lett.
\textbf{82} (1999) 4971.\\
10) M. Li: Phys. Lett. B \textbf{603} (2004) 1.\\
11) Q. G. Huang and Y. G. Gong: JCAP \textbf{0408} (2004) 006; X.
Zhang and F.-Q. Wu: Phys. Rev. D \textbf{72} (2005) 043524; ibid.
\textbf{76} (2007) 023502.\\
12) T. P Sotiriou and V. Faraoni: Rev. Mod. Phys. \textbf{82} (2010)
451; A. De Felice and S. Tsujikawa: Living Rev. Rel. \textbf{13}
(2010) 3; S. Nojiri and S. D. Odintsov: Phys. Rep. \textbf{505}
(2011) 59.\\
13) R. Ferraro and F. Fiorini: Phys. Rev. D \textbf{75} (2007)
08403; G. R. Bengochea and R. Ferraro: Phys. Rev. D \textbf{79}
(2009) 124019; E. V. Linder: Phys. Rev. D \textbf{81} (2010)
127301.\\
14) T. Harko, F. S. N. Lobo, S. Nojiri and S. D. Odintsov: Phys.
Rev. D \textbf{84} (2011) 024020.\\
15) M. Sharif and M. Zubair: JCAP \textbf{03} (2012) 028 [Erratum
ibid. \textbf{05} (2012) E01].\\
16) M. Jamil, D. Momeni, M. Raza and R.
Myrzakulov: Eur. Phys. J. C \textbf{72} (2012) 1999.\\
17) M. J. S. Houndjo and O. F. Piattella: Int. J. Mod. Phys. D
\textbf{21} (2012) 1250024.\\
18) M. J. S. Houndjo: Int. J. Mod. Phys. D \textbf{21} (2012)
1250003.\\
19) M. Sharif and M. Zubair: J. Phys. Soc. Jpn. \textbf{81} (2012)
114005.\\
20) M. Sharif and M. Zubair: J. Phys. Soc. Jpn. \textbf{82} (2013)
014002.\\
21) S. Capozziello, V. F. Cardone and A. Troisi: Phys. Rev. D
\textbf{71} (2005) 043503.\\
22) M. R. Setare: Int. J. Mod. Phys. D \textbf{17} (2008 )2219.\\
23) X. Wu and Z.-H. Zhu: Phys. Lett. B \textbf{660} (2008) 293.\\
24) C.-J. Feng: Phys. Lett. B \textbf{676} (2009) 168.\\
25) K. Karami and M. S. Khaledian: JHEP \textbf{03} (2011) 086.\\
26) M. H. Daouda, M. E. Rodrigues and M. J. S. Houndjo: Eur. Phys.
J. C \textbf{72} (2012) 1893.\\
27) S. Carloni, R. Goswami and P. K. S. Dunsby: Class. Quantum Grav.
\textbf{29} (2012) 135012.\\
28) L. D. Landau, and E. M. Lifshitz: \emph{The Classical Theory
of Fields} (Butterworth-Heinemann, 2002).\\
29) S. Nojiri, S. D. Odintsov and S. Tsujikawa: Phys. Rev. D
\textbf{71} (2005) 063004; K. Bamba, R. Myrzakulov, S. Nojiri and S.
D. Odintsov: Phys. Rev. D \textbf{85} (2012) 104036.\\
30) H. Wei and R. G. Cai: Phys. Lett. B \textbf{660} (2008) 113;
ibid. \textbf{663} (2008) 1.\\
31) J.-P. Wu, D.-Z. Ma and Y. Ling: Phys. Lett. B \textbf{663}
(2008) 152; A. Sheykhi: Phys. Rev. D \textbf{81} (2010) 023525; M.
Jamil and E. N. Saridakis: JCAP \textbf{07} (2010) 028; M. R.
Setare: Astrophys. Space Sci. \textbf{326} (2010) 27.\\
32) R. G. Cai and S. P. Kim: JHEP \textbf{02} (2005) 050; M. Akbar
and R. G. Cai: Phys. Rev. D \textbf{75} (2007) 084003.\\
33) H. M. Sadjadi: Phys. Rev. D \textbf{76} (2007) 104024.\\
34) K. Bamba and C. Q. Geng: Phys. Lett. B \textbf{679} (2009) 282;
JCAP \textbf{06} (2010) 014; ibid. \textbf{11} (2011) 008.\\
35) U. Debnath,  S. Chattopadhyay, I. Hussain, M. Jamil and R.
Myrzakulov: Eur. Phys. J. C \textbf{72} (2012) 1875.\\
36) J. D. Bekenstein: Phys. Rev. D \textbf{7} (1973) 2333.\\
37) R. M. Wald: Phys. Rev. D \textbf{48} (1993) 3427.\\
38) R. Brustein, D. Gorbonos and M. Hadad: Phys. Rev. D \textbf{79}
(2009) 044025.\\
39) E. Komatsu, J. Dunkley, R. Nolta, C. L. Bennett, B. Gold, G.
Hinshaw, N. Jarosik, D. Larson, M. Limon, L. Page, D. N. Sperge, M.
Halpern, R. S. Hill, A. Kogut, S. S. Meyer, G. S. Tucker, J. L.
Weiland, E. Wollack, and E. L. Wright:
Astrophys. J. Suppl. \textbf{180} (2009) 330.\\

\end{document}